\PassOptionsToPackage{unicode}{hyperref}
\PassOptionsToPackage{hyphens}{url}
\documentclass[
  11pt,
  letterpaper,
]{article}
\usepackage{amsmath,amssymb}
\usepackage{amsthm}
\usepackage{aliascnt}
\numberwithin{equation}{section}
\usepackage{iftex}
\ifPDFTeX
  \usepackage[T1]{fontenc}
  \usepackage[utf8]{inputenc}
  \usepackage{textcomp} 
\else 
  \usepackage{unicode-math} 
  \defaultfontfeatures{Scale=MatchLowercase}
  \defaultfontfeatures[\rmfamily]{Ligatures=TeX,Scale=1}
\fi
\usepackage{lmodern}
\usepackage[margin=1in]{geometry}
\ifPDFTeX\else
\fi
\IfFileExists{upquote.sty}{\usepackage{upquote}}{}
\IfFileExists{microtype.sty}{
  \usepackage[]{microtype}
  \UseMicrotypeSet[protrusion]{basicmath} 
}{}
\makeatletter
\@ifundefined{KOMAClassName}{
  \IfFileExists{parskip.sty}{%
    \usepackage{parskip}
  }{
    \setlength{\parindent}{0pt}
    \setlength{\parskip}{6pt plus 2pt minus 1pt}}
}{
  \KOMAoptions{parskip=half}}
\makeatother
\usepackage{xcolor}
\usepackage{longtable,booktabs,array}
\usepackage{calc} 
\usepackage{etoolbox}
\makeatletter
\patchcmd\longtable{\par}{\if@noskipsec\mbox{}\fi\par}{}{}
\makeatother
\IfFileExists{footnotehyper.sty}{\usepackage{footnotehyper}}{\usepackage{footnote}}
\makesavenoteenv{longtable}
\usepackage{graphicx}
\makeatletter
\def\maxwidth{\ifdim\Gin@nat@width>\linewidth\linewidth\else\Gin@nat@width\fi}
\def\maxheight{\ifdim\Gin@nat@height>\textheight\textheight\else\Gin@nat@height\fi}
\makeatother
\setkeys{Gin}{width=\maxwidth,height=\maxheight,keepaspectratio}
\makeatletter
\def\fps@figure{htbp}
\makeatother
\ifLuaTeX
  \usepackage{selnolig}  
\fi
\usepackage{bookmark}
\usepackage[capitalise,nameinlink,noabbrev]{cleveref}
\IfFileExists{xurl.sty}{\usepackage{xurl}}{} 
\hypersetup{
  pdftitle={Target-Aware Sequential Inference: Pooled versus Stratified Anytime-Valid Designs},
  pdfauthor={Subir Hait},
  hidelinks,
  pdfcreator={LaTeX}}

\newtheorem{theorem}{Theorem}[section]
\newaliascnt{proposition}{theorem}
\newtheorem{proposition}[proposition]{Proposition}
\aliascntresetthe{proposition}
\newaliascnt{lemma}{theorem}
\newtheorem{lemma}[lemma]{Lemma}
\aliascntresetthe{lemma}
\newaliascnt{corollary}{theorem}
\newtheorem{corollary}[corollary]{Corollary}
\aliascntresetthe{corollary}
\theoremstyle{remark}
\newaliascnt{remark}{theorem}
\newtheorem{remark}[remark]{Remark}
\aliascntresetthe{remark}

\title{Target-Aware Sequential Inference:\\
Pooled versus Stratified Anytime-Valid Designs}
\author{Subir Hait\footnote{Address correspondence to Subir Hait,
  Department of Counseling, Educational Psychology, and Special
  Education, Michigan State University, 620 Farm Lane, East Lansing, MI
  48824, USA; E-mail: \texttt{haitsubi@msu.edu}.}\\
Department of Counseling, Educational Psychology, and Special
Education\\
Michigan State University, East Lansing, Michigan, USA\\
ORCID: 0009-0004-9871-9677}
\date{}

\begin{document}
\maketitle
\clearpage
\begin{center}\textbf{Abstract}\end{center}

Sequential studies with heterogeneous strata may define the estimand under a target mixture that differs from the sampling distribution. We organize the design problem by guarantee class: one pooled confidence sequence for a single prespecified target, shared pooled inference for a prespecified target set, and stratum-resolved inference when local guarantees must remain available. For target-set pooled inference we derive minimax, gap-aware, and range-capped shared proposals and an adaptive tracking rule. A specialized augmented off-policy score removes between-stratum mean heterogeneity asymptotically, recovering Neyman allocation for efficient single-target pooled inference. Robust directional certification over a convex target set is a classical intersection--union problem and does not require an M-fold vertex split. We derive a Gaussian information lower bound and an exact finite-support characteristic information that coincides with the bounded nonparametric KL-inf limit, and state sufficient conditions for first-order attainment of the Gaussian constant. Within the local-CS architecture, regularly varying boundaries imply a general allocation law with the 2/3 exponent for root-n variance-adaptive widths. Simulations and a PromptEval replay illustrate the resulting guarantee--efficiency frontier.



\clearpage
\section{Introduction}\label{introduction}

Sequential studies frequently collect data from heterogeneous strata
while the scientific estimand is a weighted average over a target
population. The target weights may represent demographic prevalence,
deployment frequencies, policy priorities, task mixtures, or any other
prespecified population composition. In this setting, three objects
should be separated: the target distribution that defines the estimand,
the sampling policy that determines where observations are collected,
and the inferential architecture used to maintain uncertainty statements
under continuous monitoring. Conflating these objects can lead to
procedures that are efficient for one objective but inefficient for
another.

Classical stratified sampling optimizes terminal estimation under a
fixed budget, most notably through Neyman allocation (Neyman 1934; Kish
1965). Sequential monitoring changes the design problem because stopping
depends on the geometry of a time-uniform uncertainty bound rather than
only on the variance of a terminal estimator. Confidence sequences and
related always-valid/betting methods provide time-uniform guarantees
under repeated inspection and data-dependent stopping (Johari et al.
2021; Shafer 2021; Howard et al. 2021; Ramdas et al. 2023; Waudby-Smith
and Ramdas 2024; Wang and Ramdas 2025), but their widths can induce
allocation criteria that differ from fixed-budget variance minimization.
The confidence-sequence literature traces back at least to Darling and
Robbins (1967) and Robbins (1970), while classical group-sequential
designs control repeated interim looks under prespecified monitoring
schedules (Pocock 1977; O\textquotesingle Brien and Fleming 1979).
Modern e-value and asymptotic time-uniform formulations further broaden
this foundation (Vovk and Wang 2021; Waudby-Smith et al. 2024b).

Anytime-valid inference for stratified data is not itself new. Turner
and Grünwald (2023), for example, construct safe sequential tests and
confidence sequences for stratified count data and show how information
can be combined across subpopulations. Adaptive stratified sampling also
has a long history when stratum variances must be learned during data
collection (Salehi et al. 2010). The question studied here is different:
given a target-weighted estimand, which inferential architecture is
required, what allocation is optimal within that architecture, and what
is the cost of maintaining local guarantees that are not scientifically
needed? Related adaptive-allocation work in stratified Monte Carlo
sampling shows that data collected within strata can be reused to learn
variance-optimal sampling proportions (Étoré and Jourdain 2010).
Especially close is Spertus, Sridhar, and Stark (2024), who study
anytime-valid inference for a bounded stratified population mean under
predictable stratum selection. They show that a simple strategy based on
separate multiplicity-adjusted stratumwise confidence bounds is
unnecessarily conservative and develop sharper union-of-intersections
tests that combine test supermartingales across strata. Accordingly, the
local-CS aggregation used below is not presented as a state-of-the-art
global test for a single weighted mean; it is a transparent architecture
for the stronger requirement that simultaneous local intervals remain
available for reporting, reweighting, or target-set robustness.

The architecture distinction is fundamental. If inference is required
only for one prespecified target mixture, observations can be sampled
from that target and pooled into a single bounded-mean process. If
simultaneous stratum-level reporting, post-hoc reweighting, or
robustness over several target mixtures is required, separate local
confidence sequences can instead be maintained and aggregated. The
second construction buys flexibility and auditability, but it pays an
additive-width and multiplicity cost. This cost can dominate finer
choices among allocation rules.

The allocation problem is also related to sequential ranking and
selection, adaptive design selection, and best-arm identification, where
sampling effort is directed toward the alternatives or comparisons that
are most informative for a final decision (Kaufmann, Cappé, and Garivier
2016; Garivier and Kaufmann 2016; Peng et al. 2016). Our setting differs
because sampling is over strata that compose a target estimand rather
than over competing systems alone, and because the target distribution
and the inferential architecture jointly determine the relevant width
objective. AI model evaluation provides a concrete illustration:
evaluation units can be stratified by task, prompt family, language,
difficulty, safety category, or deployment context, while the target
mixture need not match the benchmark sampling distribution. The
PromptEval replay in Section 10 is used only as an empirical test bed
for this more general design problem. Sequential ranking-and-selection
procedures have long coupled sampling and stopping, from
Paulson\textquotesingle s sequential selection rule to fully sequential
indifference-zone procedures (Paulson 1964; Kim and Nelson 2001). Two
recent multi-arm formulations sharpen the connection. Imbens et al.
(2026) study ``demonstration experiments'' in which the objective is to
reject a global null by showing that at least one arm has a positive
effect, comparing a pooled promising-arm procedure with time-uniform
multiple testing of individual arms and relying on asymptotic,
central-limit-based inference. Sandoval, Waudby-Smith, and Jordan
(2026) develop nonasymptotic multi-armed betting tests with log-growth
and expected-rejection-time optimality. Those problems allocate across
competing arms to detect any positive or non-null arm; the present paper
instead targets the externally weighted average \(\Delta_Q\) and, within
the local-CS architecture, optimizes aggregate confidence-sequence
width. For robust certification over a target set the two problems are
dual: their alternative is a union over arms (some arm is non-null),
whereas ours is an intersection over target vertices (every target gap is
positive), so the corresponding nulls are an intersection and a union,
respectively.

\begin{enumerate}
\def\labelenumi{\arabic{enumi}.}
\item
  First, we organize sequential inference by guarantee class and compare
  three architectures: single-target pooled inference, target-set pooled
  inference for a prespecified family of target mixtures, and fully
  stratum-resolved local inference.
\item
  Second, within the stratum-resolved architecture we derive the
  boundary-rate allocation law for regularly varying confidence-sequence
  widths, give a finite-sample predictable plug-in empirical-Bernstein
  implementation, and establish adaptive optimal-allocation tracking
  under exploration-preserving sampling.
\item
  Third, for the target-set pooled architecture we derive a minimax shared
  proposal, the exact minimum shared importance-weight range, and a
  range-capped water-filling frontier. We then derive a gap-aware minimax
  proposal for the first-order worst-target stopping scale and an adaptive
  policy that tracks it.
\item
  Fourth, we record the design consequence of the classical
  intersection--union principle (Berger 1982) for this setting: robust
  directional certification over a convex target set lets each target
  vertex be monitored at the nominal directional level without an M-fold
  split, and we identify the pooled scores used here as stratum-level
  specializations of anytime-valid off-policy inference (Waudby-Smith et
  al. 2024a), including an augmented score that recovers Neyman allocation.
\item
  Fifth, we derive an information-theoretic characteristic time for robust
  target-set certification, both in a Gaussian benchmark with a closed-form
  optimal shared allocation and in an exact finite-support form that coincides
  with the nonparametric \(\mathrm{KL}_{\inf}\) bound for outcomes in
  \([-1,1]\). We state sufficient conditions, namely allocation tracking and an
  asymptotically sharp target-level boundary, under which the Gaussian constant
  is attained to first order.
\item
  Sixth, we extend the framework to uncertain target distributions and
  familywise best-system identification, and evaluate the resulting
  guarantee--efficiency hierarchy through simulation experiments and a
  public-data replay.
\end{enumerate}

Figure 1 summarizes the target-aware design hierarchy studied throughout
the paper.

\includegraphics[width=6.2in]{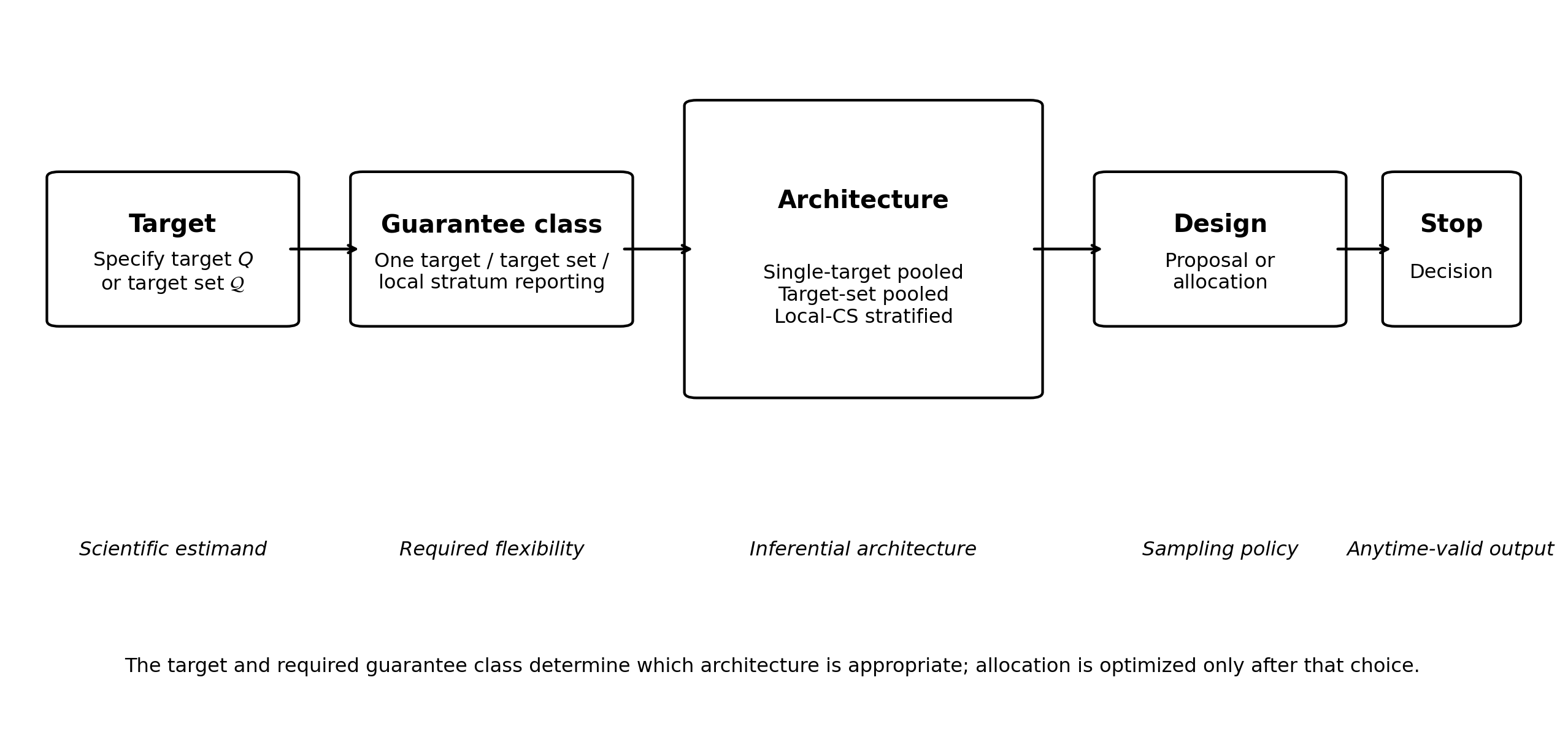}

\emph{Figure 1. Target-aware guarantee hierarchy. A single target can use one pooled confidence sequence; a prespecified target set can use simultaneous target-level pooled sequences; full local reporting requires stratum-resolved inference. Proposal or allocation is optimized only after the guarantee class and architecture are fixed.}

\section{Target-aware sequential
comparison}\label{target-aware-sequential-comparison}

\subsection{Target distribution and pairwise
gap}\label{target-distribution-and-pairwise-gap}

Let \(j \in \{ 1,\ldots,J\}\) index evaluation strata. A stratum may
represent a prompt family, task domain, language, difficulty bin, safety
category, deployment context, or any prespecified partition for which
the evaluator is willing to define a target weight. Let

\[Q = \left( q_{1},\ldots,q_{J} \right),\quad\quad q_{j} \geq 0,\quad\quad\sum_{j = 1}^{J}q_{j} = 1\]

be the target distribution. The target may come from deployment logs, a
policy-defined population, or an explicitly chosen scientific mixture.

For two systems \(a\) and \(b\), let \(D_{j,r}\) denote the paired
performance difference on the \(r\)th evaluation drawn from stratum
\(j\). We assume

\[\mathbb{E}\left( D_{j,r} \mid \mathcal{F}_{j,r - 1} \right) = \delta_{j},\]

where \(\delta_{j}\) is the stratum-specific mean difference and
\(\mathcal{F}_{j,r}\) denotes the filtration for observations within
that stratum. The target pairwise gap is

\begin{equation}\label{eq:2.1}
\Delta_{Q} = \sum_{j = 1}^{J}q_{j}\delta_{j}.
\end{equation}

A positive \(\Delta_{Q}\) means that system \(a\) is better than system
\(b\) for the target population \(Q\) under the chosen performance
metric.

The target distribution and sampling distribution need not coincide. At
global evaluation time \(t\), let \(A_{t} \in \{ 1,\ldots,J\}\) be the
selected stratum and let

\[\pi_{t}(j) = \mathbb{P}\left( A_{t} = j \mid \mathcal{H}_{t - 1} \right)\]

be the possibly adaptive proposal probability, where
\(\mathcal{H}_{t - 1}\) contains all information available after time
\(t - 1\). The proposal vector \(\pi_{t}\) is measurable with respect to
\(\mathcal{H}_{t - 1}\). After \(A_{t}\) is drawn using only this
past-measurable proposal, define the pre-outcome field

\[\mathcal{H}_{t}^{-} = \sigma\left( \mathcal{H}_{t - 1},A_{t} \right),\]

and reveal the current outcome only afterward, so
\(\mathcal{H}_{t} = \sigma\left( \mathcal{H}_{t}^{-},D_{t} \right)\).
The sampling assumption used below is

\[\mathbb{E}\left( D_{t} \mid \mathcal{H}_{t}^{-} \right) = \delta_{A_{t}}.\]

This separation is crucial: \(Q\) defines what is being estimated;
\(\pi_{t}\) determines where new information is collected.

Throughout, one \emph{evaluation unit} means one sampled condition in a
stratum with the systems needed for the stated comparison evaluated on
that condition. Thus a pairwise unit typically entails two model calls,
whereas a \(K\)-system unit entails up to \(K\) model calls if all
systems are evaluated. The cost \(c_{j}\) may absorb model-specific
token, latency, or monetary costs. Reporting evaluation units keeps the
statistical allocation distinct from a particular API accounting
convention.

\section{Pooled target-specific
inference}\label{pooled-target-specific-inference}

The stratified construction is not necessary when the sole inferential
target is one prespecified \(Q\) and simultaneous stratum-level
intervals are not required. There is then a simpler route. Suppose at
global time \(t\) the evaluator draws \(A_{t}\) from a predictable
proposal \(\pi_{t}\) with \(\pi_{t}(j) > 0\) whenever \(q_{j} > 0\),
then observes a paired difference \(D_{t}\) from the selected stratum.
Define

\begin{equation}\label{eq:3.1}
Z_{t} = \frac{q_{A_{t}}}{\pi_{t}\left( A_{t} \right)}D_{t}.
\end{equation}

By iterated expectation,

\begin{equation}\label{eq:3.2}
\mathbb{E}\left( Z_{t} \mid \mathcal{H}_{t - 1} \right) = \sum_{j}^{}\pi_{t}(j)\frac{q_{j}}{\pi_{t}(j)}\delta_{j} = \Delta_{Q}.
\end{equation}

Thus \(\left( Z_{t} - \Delta_{Q} \right)\) is a martingale-difference
sequence, and an off-policy confidence sequence can be applied directly
to the pooled stream (Karampatziakis, Mineiro, and Ramdas 2021). This
avoids a union bound over the \(J\) strata. More generally, inference
after adaptive experimentation must explicitly account for
data-dependent sampling or weighting; Hadad et al. (2021) provide a
complementary large-sample treatment for adaptively collected
policy-evaluation data.

The most transparent special case is \emph{target sampling}: set
\(\pi_{t} = Q\) for every \(t\). Then

\begin{equation}\label{eq:3.3}
Z_{t} = D_{t} \in \lbrack - 1,1\rbrack,
\end{equation}

so a single bounded-mean PrPl-EB sequence at level \(\alpha\) targets
\(\Delta_{Q}\) without importance weights and without the \(\alpha/J\)
local split. For one fixed target, this is a strong default benchmark
against which any stratum-resolved construction should be compared.

For a fixed proposal \(\pi\), write
\(m_{2j} = \mathbb{E}\left( D_{j}^{2} \right) = \sigma_{j}^{2} + \delta_{j}^{2}\).
The direct estimator's variance is

\begin{equation}\label{eq:3.4}
Var\left( Z_{t} \right) = \sum_{j}^{}\frac{q_{j}^{2}m_{2j}}{\pi_{j}} - \Delta_{Q}^{2}.
\end{equation}

Ignoring finite-time range effects, the proposal minimizing the first
term is

\begin{equation}\label{eq:3.5}
\pi_{j}^{dir,var} \propto q_{j}\sqrt{m_{2j}} = q_{j}\sqrt{\sigma_{j}^{2} + \delta_{j}^{2}}.
\end{equation}

This is not exactly the stratified Neyman allocation
\(q_{j}\sigma_{j}\): randomizing the stratum adds the between-stratum
second-moment term \(\delta_{j}^{2}\). More importantly for confidence
sequences, Equation~\eqref{eq:3.5} need not be finite-time optimal because the
importance-weight range

\[B(\pi) = \max_{j}q_{j}/\pi_{j}\]

also enters bounded sequential inference. A proposal that is
variance-efficient but too spiky can therefore widen a finite-time
confidence sequence, paralleling the querying-distribution phenomenon
observed by Hsu and Shekhar (2026).

\begin{proposition}[Architecture tradeoff and first-order penalty]\label{prop:architecture} (a) For a single prespecified target Q, direct target
sampling produces one bounded pooled stream whose conditional mean is
the target gap. Any valid bounded-mean confidence sequence therefore
gives anytime-valid target inference without a multiplicity factor over
strata. A stratum-resolved construction instead maintains simultaneous
local intervals, supporting local reporting, post-hoc deterministic
reweighting, and robust inference over target sets, at the cost of
additive local widths and simultaneous coverage.

(b) Consider a symmetric root-n benchmark with J equally weighted
strata, equal costs, and local first-order half-width
\(r(\eta)/\sqrt{n}\). If the pooled target-sampling process has
first-order half-width \(r(\alpha)/\sqrt{N}\), then the optimized
stratified construction allocates N/J observations to each stratum and
has first-order half-width \(r(\alpha/J)\sqrt{J/N}\), whereas the pooled
half-width is \(r(\alpha)/\sqrt{N}\). Their ratio is therefore
\(\sqrt{J}\, r(\alpha/J)/r(\alpha)\). For standard boundaries
\(r(\alpha/J) \geq r(\alpha)\), so the width penalty is at least
\(\sqrt{J}\); for a fixed nonzero target gap, the corresponding
first-order stopping-cost penalty is order J.

\end{proposition}

The calculation follows because symmetry makes N/J the width-minimizing
local allocation and the weighted sum of J local half-widths preserves
the \(\sqrt{J}\) factor. The benchmark is intentionally stylized:
heterogeneity in stratum means, variances, costs, or importance-weight
ranges changes the constant. Its role is to make explicit why
architecture can dominate allocation as the number of required local
guarantees grows. Pooled inference is preferred when only one target
comparison is required; stratified inference is warranted when local
auditability or target flexibility is itself part of the scientific
requirement.

\section{Stratified sequential design
objectives}\label{stratified-sequential-design-objectives}

\subsection{Fixed-budget estimation}\label{fixed-budget-estimation}

Suppose \(n_{j}\) independent paired evaluations are collected in
stratum \(j\), with

\[Var\left( D_{j,r} \right) = \sigma_{j}^{2},\]

and define

\[{\widehat{\delta}}_{j} = \frac{1}{n_{j}}\sum_{r = 1}^{n_{j}}D_{j,r},\quad\quad{\widehat{\Delta}}_{Q} = \sum_{j = 1}^{J}q_{j}{\widehat{\delta}}_{j}.\]

Under independent strata,

\begin{equation}\label{eq:4.1}
Var\left( {\widehat{\Delta}}_{Q} \right) = \sum_{j = 1}^{J}\frac{q_{j}^{2}\sigma_{j}^{2}}{n_{j}}.
\end{equation}

Let one observation in stratum \(j\) cost \(c_{j} > 0\) and impose

\[\sum_{j = 1}^{J}c_{j}n_{j} \leq B.\]

The classical variance-minimizing allocation is the cost-adjusted Neyman
rule (Neyman 1934).

\begin{proposition}[Estimation-optimal allocation]\label{prop:neyman} Under
Equation~\eqref{eq:4.1}, treating \(n_{j} > 0\) as continuous, the allocation
minimizing \(Var\left( {\widehat{\Delta}}_{Q} \right)\) subject to
\(\sum_{j}^{}c_{j}n_{j} = B\) is

\begin{equation}\label{eq:4.2}
n_{j}^{est} = \frac{B\, q_{j}\sigma_{j}/\sqrt{c_{j}}}{\sum_{\ell = 1}^{J}q_{\ell}\sigma_{\ell}\sqrt{c_{\ell}}}.
\end{equation}

The minimum variance is

\begin{equation}\label{eq:4.3}
V_{\min} = \frac{\left( \sum_{j}^{}q_{j}\sigma_{j}\sqrt{c_{j}} \right)^{2}}{B}.
\end{equation}

\end{proposition}

A derivation is given in Appendix~A. This is standard
stratified-sampling theory, not a new result. It serves as the reference
point for the sequential objective below. Note that if \(\sigma_{j}\)
and \(c_{j}\) are constant, Equation~\eqref{eq:4.2} reduces to sampling
proportional to \(q_{j}\). If \(Q\) is known in advance, this avoids the
effective-sample-size loss caused by sampling uniformly and reweighting
afterward. Kish-type effective sample-size calculations (Kish 1965)
remain useful diagnostics for already-collected weighted samples, but
they are not the design solution when prospective allocation is
possible.

\subsection{Sequential identification with simultaneous
stratum-level confidence
sequences}\label{sequential-identification-with-simultaneous-stratum-level-confidence-sequences}

This subsection assumes that the evaluator wants simultaneous
stratum-level inference---for example, because subject-specific results
must be reported, because \(Q\) may be changed after data collection, or
because a family of target distributions will be queried from the same
local confidence event. A sequential procedure stops when the
uncertainty interval for \(\Delta_{Q}\) excludes zero. Suppose stratum
\(j\) has a valid local confidence sequence with half-width \(b_{j}(n)\)
after \(n\) local observations. A conservative target-level confidence
sequence is obtained by weighting the local intervals, with half-width

\begin{equation}\label{eq:4.4}
W\left( \mathbf{n} \right) = \sum_{j = 1}^{J}q_{j}b_{j}\left( n_{j} \right).
\end{equation}

Within this stratum-resolved architecture, the allocation that makes
\(W\) small need not minimize Equation~\eqref{eq:4.1}. The conclusion is
conditional on using additive aggregation of simultaneous local
intervals; Section~3 gives a pooled target-specific alternative for
which this geometry disappears.

We state the allocation law for a broad class of regularly varying
widths. Suppose, for fixed \(J\) and active strata with \(q_{j} > 0\),

\begin{equation}\label{eq:4.5}
b_{j}(n) = r_{j}L(n)n^{- \beta}\{ 1 + o(1)\},\quad\quad r_{j} > 0,\quad\beta > 0,
\end{equation}

where \(L\) is a positive slowly varying function common across strata
(or the stratum-specific slowly varying factors are asymptotically
equivalent). This common-\(\beta\) representation is a first-order model
for nondegenerate active strata. Near-degenerate bounded strata can
exhibit faster empirical-Bernstein shrinkage and should be handled
separately or by exact finite-time numerical optimization rather than
forced into the common-rate formula. For allocations with \(n_{j}/B\)
bounded away from zero, slow variation implies
\(L\left( n_{j} \right)/L(B) \rightarrow 1\). After removing the common
factor \(L(B)B^{- \beta}\), the asymptotic target-width problem is
therefore

\[{\widetilde{W}}_{\beta}\left( \mathbf{n} \right) = \sum_{j = 1}^{J}a_{j}n_{j}^{- \beta},\quad\quad a_{j} = q_{j}r_{j}.\]

\begin{proposition}[Stratified boundary-rate-optimal allocation]\label{prop:boundary-rate}
Under Equation~\eqref{eq:4.5}, the asymptotically width-minimizing allocation
shares, treating \(n_{j} > 0\) as continuous, are

\begin{equation}\label{eq:4.6}
n_{j}^{id} = \frac{B\, a_{j}^{1/(1 + \beta)}c_{j}^{- 1/(1 + \beta)}}{\sum_{\ell = 1}^{J}a_{\ell}^{1/(1 + \beta)}c_{\ell}^{\beta/(1 + \beta)}}.
\end{equation}

Writing

\[S_{\beta} = \sum_{j = 1}^{J}a_{j}^{1/(1 + \beta)}c_{j}^{\beta/(1 + \beta)},\]

the minimum target half-width satisfies

\begin{equation}\label{eq:4.7}
W_{\min}(B) = L(B)\frac{S_{\beta}^{1 + \beta}}{B^{\beta}}\{ 1 + o(1)\}.
\end{equation}

For the root-\(n\) case \(\beta = 1/2\),

\begin{equation}\label{eq:4.8}
n_{j}^{id} = \frac{B\, a_{j}^{2/3}c_{j}^{- 2/3}}{\sum_{\ell = 1}^{J}a_{\ell}^{2/3}c_{\ell}^{1/3}},
\end{equation}

and the minimum of the root-\(n\) design proxy is

\begin{equation}\label{eq:4.9}
{\widetilde{W}}_{\min} = \frac{\left( \sum_{j}^{}a_{j}^{2/3}c_{j}^{1/3} \right)^{3/2}}{\sqrt{B}}.
\end{equation}

With equal costs and a range-only local sequence whose leading constant
is common across strata, \(r_{j} = r\), so

\begin{equation}\label{eq:4.10}
n_{j}^{id} \propto q_{j}^{2/3}.
\end{equation}

For a variance-adaptive local sequence with leading constant
proportional to \(\sigma_{j}\), the analogous first-order rule is
approximately

\begin{equation}\label{eq:4.11}
n_{j}^{id} \propto \left( q_{j}\sigma_{j} \right)^{2/3}.
\end{equation}

\end{proposition}

A derivation is given in Appendix~B. Neither rule is the Neyman
allocation \(q_{j}\sigma_{j}\). This difference is a property of the
stratum-resolved additive-width construction, not a universal separation
between estimation and sequential identification.

\begin{corollary}[Target-proportional sampling reversal]\label{cor:reversal} Assume
equal costs. For a common-constant root-\(n\) range boundary, uniform
allocation \(n_{j} = B/J\) has leading width

\[W_{U} = \frac{r\sqrt{J}}{\sqrt{B}},\]

whereas target-proportional allocation \(n_{j} = Bq_{j}\) has

\[W_{Q} = \frac{r}{\sqrt{B}}\sum_{j = 1}^{J}\sqrt{q_{j}} \leq W_{U},\]

with equality if and only if \(Q\) is uniform. By contrast, their
fixed-budget estimator variances are

\begin{equation}\label{eq:4.12}
V_{U} = \frac{J}{B}\sum_{j}^{}q_{j}^{2}\sigma_{j}^{2},\quad\quad V_{Q} = \frac{1}{B}\sum_{j}^{}q_{j}\sigma_{j}^{2},
\end{equation}

so \(Q\)-proportional sampling improves fixed-budget variance over
uniform sampling if and only if

\begin{equation}\label{eq:4.13}
\sum_{j}^{}q_{j}\sigma_{j}^{2} < J\sum_{j}^{}q_{j}^{2}\sigma_{j}^{2}.
\end{equation}

\end{corollary}

Thus target-proportional allocation weakly improves the leading
range-only identification width but can either improve or worsen
fixed-budget estimation variance under heteroscedasticity.

\begin{remark}[Finite-sample boundaries]\label{rem:finite-sample} Equations~\eqref{eq:4.6}--\eqref{eq:4.11}
describe asymptotic allocation shares after slowly varying factors are
separated. Exact mixture or betting boundaries contain logarithmic and
finite-time terms. For a fixed confidence-sequence construction, a
practitioner can therefore solve

\[\min_{n_{1},\ldots,n_{J}}\sum_{j}^{}q_{j}b_{j}\left( n_{j} \right)\quad\text{subject to}\quad\sum_{j}^{}c_{j}n_{j} \leq B\]

numerically. The analytic allocation law is a transparent first-order
design principle and a useful initialization, not a claim of exact
finite-sample optimality for every boundary.

\end{remark}

\subsection{What the stratified reversal does---and does
not---mean}\label{what-the-stratified-reversal-doesand-does-notmean}

\cref{cor:reversal} is useful, but its scope must be explicit. It compares
fixed-budget stratified estimation with a sequential procedure that
requires simultaneous local confidence sequences and then aggregates
their widths in \(L_{1}\) form. The reversal is therefore a design
result \emph{within that guarantee structure}. It does not imply that
every anytime-valid procedure for \(\Delta_{Q}\) has a different optimal
allocation from fixed-budget estimation. In particular, if only one
target \(Q\) is prespecified, a pooled construction can target
\(\Delta_{Q}\) directly and avoid the additive local-width geometry
entirely. That alternative was developed in Section~3. The broader
lesson is that confidence-sequence architecture, multiplicity, and
reporting requirements can dominate the finer choice among allocation
rules.

\section{Uncertain target distributions and robust
design}\label{uncertain-target-distributions-and-robust-design}

A single target distribution may itself be uncertain. Deployment logs
can be noisy, user populations can shift, and policy makers may wish to
protect performance across several plausible mixtures. Let

\[\mathcal{Q} = conv\{ Q^{(1)},\ldots,Q^{(M)}\},\quad\quad Q^{(m)} = \left( q_{1}^{(m)},\ldots,q_{J}^{(m)} \right).\]

For the boundary-rate proxy of \cref{prop:boundary-rate}, define

\[f_{m}\left( \mathbf{n} \right) = \sum_{j = 1}^{J}q_{j}^{(m)}r_{j}n_{j}^{- \beta}.\]

A robust evaluation design solves

\begin{equation}\label{eq:5.1}
\min_{\mathbf{n} > 0:\,\sum_{j}^{}c_{j}n_{j} = B}\ \max_{1 \leq m \leq M}f_{m}\left( \mathbf{n} \right).
\end{equation}

Because each \(f_{m}\) is convex in the positive allocation vector, the
epigraph formulation is a convex program.

\begin{proposition}[Least-favourable target mixture]\label{prop:least-favourable} For
problem~\eqref{eq:5.1}, there exists a vector
\(\lambda = \left( \lambda_{1},\ldots,\lambda_{M} \right)\) on the
probability simplex, with positive mass only on worst-case target
scenarios active at the optimum, such that

\[{\bar{q}}_{j} = \sum_{m = 1}^{M}\lambda_{m}q_{j}^{(m)}.\]

Assume \({\bar{q}}_{j}r_{j} > 0\) for every retained stratum; strata
with zero coefficient may be removed from the optimization or assigned
only an externally imposed exploration floor. Then the robust allocation
is

\begin{equation}\label{eq:5.2}
n_{j}^{rob} = \frac{B\left( {\bar{q}}_{j}r_{j} \right)^{1/(1 + \beta)}c_{j}^{- 1/(1 + \beta)}}{\sum_{\ell}^{}\left( {\bar{q}}_{\ell}r_{\ell} \right)^{1/(1 + \beta)}c_{\ell}^{\beta/(1 + \beta)}}.
\end{equation}

Thus target uncertainty changes the ordinary identification design only
through a least-favourable mixture \(\bar{Q}\). For root-\(n\)
variance-sensitive boundaries and equal costs,

\[n_{j}^{rob} \propto \left( {\bar{q}}_{j}\sigma_{j} \right)^{2/3}.\]

\end{proposition}

A derivation is given in Appendix~F. The result is useful
computationally as well as conceptually. Rather than searching over all
allocations and target mixtures simultaneously, one may optimize over
the low-dimensional simplex of candidate target scenarios and then apply
the closed-form allocation corresponding to the resulting \(\bar{Q}\).

\section{Anytime-valid inference under adaptive stratum
sampling}\label{anytime-valid-inference-under-adaptive-stratum-sampling}

\subsection{Bounded-mean confidence sequences: range-only and
variance-adaptive}\label{bounded-mean-confidence-sequences-range-only-and-variance-adaptive}

Assume paired differences satisfy \(D_{j,r} \in \lbrack - 1,1\rbrack\).
Hoeffding's lemma makes \(D_{j,r} - \delta_{j}\) conditionally
1-sub-Gaussian (Hoeffding 1963). A normal-mixture e-process yields a
transparent two-sided local confidence sequence. For tuning parameter
\(\rho > 0\) and local error level \(\alpha_{0}\), define

\begin{equation}\label{eq:6.1}
b_{\rho,\alpha_{0}}(n) = \frac{\sqrt{(n + \rho)\log\{(n + \rho)/\left( \rho\alpha_{0}^{2} \right)\}}}{n}.
\end{equation}

Then

\[C_{j,n}^{range} = \left\lbrack {\overline{D}}_{j,n} - b_{\rho,\alpha_{0}}(n),{\overline{D}}_{j,n} + b_{\rho,\alpha_{0}}(n) \right\rbrack \cap \lbrack - 1,1\rbrack\]

is a valid local sequence. This normal-mixture-style construction
follows classical mixture-martingale ideas (Robbins 1970; Howard et al.
2021) and is deliberately conservative because it uses only the support
range. This lineage includes early confidence sequences for means and
related parameters (Darling and Robbins 1967) and the modern e-value
formulation of nonnegative evidence processes (Vovk and Wang 2021).

For a practical variance-adaptive sequence we use the predictable
plug-in empirical-Bernstein (PrPl-EB) construction of Waudby-Smith and
Ramdas (2024). Transform \(D_{i}\) to
\(X_{i} = \left( D_{i} + 1 \right)/2\) in \(\lbrack 0,1\rbrack\). Let
\({\widehat{\mu}}_{i - 1}\) be any predictable estimate of the bounded
mean, choose a predictable betting fraction \(\lambda_{i}\) strictly
between 0 and a fixed cap \(c < 1\), and define the empirical-Bernstein
variance term below. This construction sits in a longer
empirical-Bernstein tradition in which data-dependent variance
information sharpens concentration for bounded observations (Maurer and
Pontil 2009).

\[v_{i} = 4\left( X_{i} - {\widehat{\mu}}_{i - 1} \right)^{2},\quad\quad\psi_{E}(\lambda) = \frac{- \log(1 - \lambda) - \lambda}{4}.\]

Writing \(\Lambda_{n} = \sum_{i = 1}^{n}\lambda_{i}\), the closed-form
two-sided interval before running intersection is

\begin{equation}\label{eq:6.2}
C_{n}^{X,EB} = \left\lbrack \frac{\sum_{i = 1}^{n}\lambda_{i}X_{i}}{\Lambda_{n}}\  \pm \ \frac{\log\left( 2/\alpha_{0} \right) + \sum_{i = 1}^{n}v_{i}\psi_{E}\left( \lambda_{i} \right)}{\Lambda_{n}} \right\rbrack \cap \lbrack 0,1\rbrack.
\end{equation}

Its running intersection is again a \(\left( 1 - \alpha_{0} \right)\)
confidence sequence. Mapping the endpoints by \(d = 2x - 1\) gives a
valid confidence sequence for \(\delta_{j}\).

We use the fully predictable variance-adaptive choice

\begin{equation}\label{eq:6.3}
\lambda_{i} = \min\left\{ c,\sqrt{\frac{2\log\left( 2/\alpha_{0} \right)}{{\widetilde{\sigma}}_{i - 1}^{2}\, i\log(1 + i)}} \right),
\end{equation}

where \({\widetilde{\sigma}}_{i - 1}^{2}\) is a positive regularized
estimate of the variance of \(X\) based only on past observations.
Validity of Equation~\eqref{eq:6.2} requires predictability of \(\lambda_{i}\),
not correctness of the variance estimate. The variance estimate affects
efficiency. Waudby-Smith and Ramdas (2024) show that PrPl-EB adapts to
the unknown variance and approaches the ideal Bernstein width based on
the true variance asymptotically. Consequently, under i.i.d. stratum
sampling its leading local width coefficient is proportional to
\(\sigma_{j}\), justifying the \(\left( q_{j}\sigma_{j} \right)^{2/3}\)
identification design derived earlier. Asymptotic confidence sequences
based on time-uniform central limit theory provide a complementary route
when exact nonasymptotic bounded-data guarantees are not required
(Waudby-Smith et al. 2024b).

\subsection{Predictable adaptive allocation and local
time}\label{predictable-adaptive-allocation-and-local-time}

The global procedure may choose strata adaptively. With the pre-outcome
filtration defined above, let

\[N_{j}(t) = \sum_{s = 1}^{t}\mathbf{1}\left( A_{s} = j \right).\]

Let

\[\tau_{j,r} = \inf\{ t:N_{j}(t) = r\}\]

be the global time of the \(r\)th visit to stratum \(j\), with
\(\tau_{j,r} = \infty\) if that visit never occurs. Let
\(X_{j,r} = D_{\tau_{j,r}}\) on \(\{\tau_{j,r} < \infty\}\).

\begin{lemma}[Predictable local-time validity]\label{lem:local-time} Suppose
\(\pi_{t}\) is \(\mathcal{H}_{t - 1}\)-measurable, \(A_{t}\) is drawn
before observing \(D_{t}\), and

\[\mathbb{E}\left( D_{t} \mid \mathcal{H}_{t}^{-} \right) = \delta_{A_{t}}.\]

Let \(\left( M_{j,r}\left( \delta_{j} \right) \right)_{r \geq 0}\) be a
nonnegative local test supermartingale constructed from the successive
observations in stratum \(j\) under the null mean \(\delta_{j}\). Then
the time-changed process

\[{\widetilde{M}}_{j,t}\left( \delta_{j} \right) = M_{j,N_{j}(t)}\left( \delta_{j} \right)\]

is a nonnegative supermartingale with respect to the global filtration.
If stratum \(j\) is visited only finitely often, the process is held
constant after its last visit. Consequently,

\[\mathbb{P}\left( \exists t \geq 0:\ {\widetilde{M}}_{j,t}\left( \delta_{j} \right) \geq 1/\alpha_{0} \right) \leq \alpha_{0}.\]

\end{lemma}

A proof is given in Appendix~C. The lemma is the technical reason
adaptive allocation does not invalidate stratum-wise confidence
sequences. The proposal may depend on the entire observed past, but the
current outcome cannot be used to choose the stratum on which that same
outcome is observed. The broader adaptive-experimentation literature
likewise emphasizes predictable or data-dependent allocation together
with inferential adjustments that account for the sampling rule (Hadad
et al. 2021).

\subsection{Learning the variance-optimal identification
allocation}\label{learning-the-variance-optimal-identification-allocation}

The first-order variance-optimal allocation depends on unknown
\(\sigma_{j}\). It can nevertheless be learned online without
sacrificing validity. For clarity, consider equal per-unit costs and
\(q_{j} > 0\) for all target-relevant strata. After a positive pilot
sample, let \({\widehat{\sigma}}_{j,t}\) be the sample standard
deviation within stratum \(j\) based on observations available by global
time \(t\), stabilized by an arbitrarily small positive floor. Define

\begin{equation}\label{eq:6.4}
{\widehat{w}}_{j,t} = \frac{\left( q_{j}{\widehat{\sigma}}_{j,t} \right)^{2/3}}{\sum_{\ell = 1}^{J}\left( q_{\ell}{\widehat{\sigma}}_{\ell,t} \right)^{2/3}},\quad\quad\pi_{t + 1}(j) = \left( 1 - \varepsilon_{t} \right){\widehat{w}}_{j,t} + \varepsilon_{t}\nu_{j},
\end{equation}

where \(\nu_{j} > 0\), \(\sum_{j}^{}\nu_{j} = 1\),
\(\varepsilon_{t} \downarrow 0\), and
\(\sum_{t}^{}\varepsilon_{t} = \infty\).

\begin{theorem}[Adaptive optimal-allocation tracking]\label{thm:tracking} Suppose the
observations within each stratum are i.i.d., bounded, and have
\(0 < \sigma_{j} < \infty\). Under the predictable allocation
rule~\eqref{eq:6.4},

\begin{equation}\label{eq:6.5}
\frac{N_{j}(t)}{t} \rightarrow w_{j}^{opt} = \frac{\left( q_{j}\sigma_{j} \right)^{2/3}}{\sum_{\ell = 1}^{J}\left( q_{\ell}\sigma_{\ell} \right)^{2/3}},\quad\quad j = 1,\ldots,J,\quad\text{a.s.}
\end{equation}

\end{theorem}

A proof is given in Appendix~E. The exploration term has two roles. It
guarantees infinitely many observations in every target-relevant
stratum, so the variance estimates are strongly consistent, and it
prevents the adaptive policy from becoming trapped by early noise.
Because \(\varepsilon_{t} \rightarrow 0\), the exploration cost vanishes
as a fraction of the total budget. This tracking perspective is closely
related to adaptive optimal allocation in stratified Monte Carlo
sampling, where estimated stratum variability is used to drive
convergence toward variance-optimal proportions (Étoré and Jourdain
2010).

\begin{corollary}[First-order width efficiency]\label{cor:width-efficiency} Assume in
addition that the local variance-adaptive confidence-sequence
half-widths obey

\[b_{j}(n) = \kappa\sigma_{j}L(n)n^{- 1/2}\{ 1 + o(1)\}\]

almost surely, where \(L\) is a common slowly varying function and
\(\kappa > 0\). Let \(W_{t} = \sum_{j}^{}q_{j}b_{j}\{ N_{j}(t)\}\). Then

\begin{equation}\label{eq:6.6}
\frac{W_{t}}{W_{t}^{opt}}\overset{a.s.}{\rightarrow}1,
\end{equation}

where \(W_{t}^{opt}\) is the first-order minimum obtained by allocating
\(t\) observations according to the optimal shares \(w^{opt}\).

\end{corollary}

\cref{cor:width-efficiency} is deliberately a width statement. Allocation convergence
alone does not imply a universal ratio-optimality statement for random
stopping times without an additional asymptotic regime, such as
shrinking target gaps or \(\alpha \downarrow 0\).

\subsection{Target-level confidence
sequence}\label{target-level-confidence-sequence}

Suppose each local sequence
\(C_{j,N_{j}(t)} = \left\lbrack L_{j,t},U_{j,t} \right\rbrack\) has
simultaneous coverage at level \(1 - \alpha/J\). Define

\begin{equation}\label{eq:6.7}
L_{Q,t} = \sum_{j = 1}^{J}q_{j}L_{j,t},\quad\quad U_{Q,t} = \sum_{j = 1}^{J}q_{j}U_{j,t}.
\end{equation}

\begin{theorem}[Anytime-valid target interval]\label{thm:target-interval} Under \cref{lem:local-time},
if the \(J\) local confidence sequences each have time-uniform error
probability at most \(\alpha/J\), then

\[\mathbb{P}\{\Delta_{Q} \in \left\lbrack L_{Q,t},U_{Q,t} \right\rbrack\text{ for every }t\} \geq 1 - \alpha.\]

\end{theorem}

This statement remains valid under any predictable adaptive allocation
rule. For the interval width to converge to zero by the stratum-wise
construction, every stratum with \(q_{j} > 0\) must receive infinitely
many observations.

The last condition explains why an exploration floor is not cosmetic. A
policy that permanently starves a positive-weight target stratum cannot
identify the sign of the target mean with the stratum-wise construction
because that stratum's contribution to the aggregate width never
vanishes.

\subsection{Anytime-valid inference uniformly over target
uncertainty}\label{anytime-valid-inference-uniformly-over-target-uncertainty}

The same joint stratum event also supports inference for every target in
\(\mathcal{Q}\) without a second union bound over the \(M\) target
scenarios. Define

\begin{equation}\label{eq:6.8}
L_{\mathcal{Q},t} = \min_{1 \leq m \leq M}\sum_{j}^{}q_{j}^{(m)}L_{j,t},\quad\quad U_{\mathcal{Q},t} = \max_{1 \leq m \leq M}\sum_{j}^{}q_{j}^{(m)}U_{j,t}.
\end{equation}

\begin{proposition}[Uniform target-set confidence sequence]\label{prop:uniform-targetset} On
the joint local coverage event used in \cref{thm:target-interval},

\[\Delta_{Q} \in \left\lbrack L_{\mathcal{Q},t},U_{\mathcal{Q},t} \right\rbrack\quad\text{for every }Q \in \mathcal{Q}\text{ and every }t.\]

Hence the interval in Equation~\eqref{eq:6.8} has time-uniform coverage at least
\(1 - \alpha\) simultaneously over all \(Q \in \mathcal{Q}\), with no
additional \(M\)-fold multiplicity factor.

\end{proposition}

A proof is given in Appendix~D. The reason is geometric rather than
probabilistic: conditional on simultaneous coverage of all stratum
means, both extrema of a linear functional over a convex hull occur at
vertices. Robust target uncertainty therefore reuses the same joint
stratum-wise confidence event.

\subsection{Direct off-policy confidence
sequences}\label{direct-off-policy-confidence-sequences}

There is an alternative that avoids splitting error probability over
strata. This construction is a particularly simple off-policy-evaluation
special case. More general anytime-valid off-policy methods for contextual
bandits allow adaptive logging policies, doubly robust scores, empirical
variance adaptation, and need not require a known uniform bound on importance
weights (Waudby-Smith et al. 2024a). Here the action is simply the selected
stratum, the logging propensity is the known design probability, and the target
policy is the externally specified target mixture. If the proposal probabilities
are known and satisfy positivity,

\[\pi_{t}(j) > 0\quad\text{whenever }q_{j} > 0,\]

then

\begin{equation}\label{eq:6.9}
Z_{t} = \frac{q_{A_{t}}}{\pi_{t}\left( A_{t} \right)}D_{t}
\end{equation}

satisfies
\(\mathbb{E}\left( Z_{t} \mid \mathcal{H}_{t - 1} \right) = \Delta_{Q}\).
Off-policy confidence-sequence methods can therefore target
\(\Delta_{Q}\) directly and avoid a \(J\)-fold union bound
(Karampatziakis, Mineiro, and Ramdas 2021). This can be substantially
more efficient.

The tradeoff is structural. Direct importance weighting requires
known/logged propensities and can suffer from large weights when
\(\pi_{t}(j)\) is small relative to \(q_{j}\). Clipping improves
stability but changes the exact unbiasedness argument unless bias is
handled explicitly. The stratum-wise construction does not require
propensity logging and remains valid under arbitrary predictable
allocation, but it pays a multiplicity cost and requires continuing
exploration of all target strata. We view the two approaches as
complementary rather than claiming one dominates universally.

\subsection{Target-set pooled anytime-valid inference}\label{target-set-pooled}

The two architectures above leave a useful middle ground. Suppose the
scientific protocol prespecifies a finite family of target mixtures
\(Q^{(1)},\ldots,Q^{(M)}\), with convex hull \(\mathcal Q\), but does
not require confidence sequences for every individual stratum. Under a
fixed proposal \(\pi\) satisfying \(\pi_j>0\) whenever
\(\max_m q_j^{(m)}>0\), define for each target vertex

\begin{equation}\label{eq:6.10}
Z_t^{(m)}=\frac{q_{A_t}^{(m)}}{\pi_{A_t}}D_t,\qquad
\mathbb E\!\left(Z_t^{(m)}\mid\mathcal H_{t-1}\right)=\Delta_{Q^{(m)}}.
\end{equation}

Let \(C_t^{(m)}=[L_t^{(m)},U_t^{(m)}]\) be any anytime-valid confidence
sequence for \(\Delta_{Q^{(m)}}\) at error level \(\alpha_m\), where
\(\sum_m\alpha_m\leq\alpha\). The same sampled observation therefore
updates every target-level process, but no stratumwise confidence
sequence is maintained.

\begin{proposition}[Simultaneous target-set pooled validity]\label{prop:targetset-pooled}
With probability at least \(1-\alpha\),
\[
\Delta_{Q^{(m)}}\in C_t^{(m)}\quad\text{for every }m=1,\ldots,M
\text{ and every }t.
\]
On this event, for any \(Q=\sum_m\omega_mQ^{(m)}\in\mathcal Q\),
\[
\sum_m\omega_mL_t^{(m)}\leq\Delta_Q\leq
\sum_m\omega_mU_t^{(m)}\quad\text{for every }t.
\]
In particular, \(\min_m L_t^{(m)}>0\) certifies a positive effect
uniformly over the entire convex hull, and \(\max_m U_t^{(m)}<0\)
certifies a negative effect uniformly over it.

\end{proposition}

The proposition follows from simultaneous vertex coverage and linearity
of \(Q\mapsto\Delta_Q\); a proof is given in Appendix~G. This
architecture provides prespecified target flexibility without purchasing
the stronger guarantee of simultaneous local stratum inference. It is
therefore intermediate between one-target pooled inference and the
local-CS architecture of Sections~6.4--6.5.

\subsection{Minimax and range-capped shared proposals}\label{shared-proposal}

A target-set pooled procedure uses one sampling proposal for all target
vertices. Let \(m_{2j}=\mathbb E(D_j^2)\) and, for a fixed proposal
\(\pi\), define the target-specific importance-weighted second-moment
proxy
\begin{equation}\label{eq:6.11}
S_m(\pi)=\sum_{j=1}^J\frac{(q_j^{(m)})^2m_{2j}}{\pi_j}.
\end{equation}
The shared second-moment design solves
\begin{equation}\label{eq:6.12}
\min_{\pi\in\Delta_J}\max_{1\leq m\leq M}S_m(\pi).
\end{equation}

\begin{proposition}[Minimax shared pooled proposal]\label{prop:minimax-proposal}
Assume every target-relevant stratum has \(m_{2j}>0\). There exists a
least-favourable vector \(\lambda^*\in\Delta_M\), supported only on
targets active at the optimum, such that the solution of Equation~\eqref{eq:6.12}
is
\begin{equation}\label{eq:6.13}
\pi_j^*=\frac{\sqrt{m_{2j}\sum_{m=1}^M\lambda_m^*(q_j^{(m)})^2}}
{\sum_{\ell=1}^J\sqrt{m_{2\ell}\sum_{m=1}^M\lambda_m^*(q_{\ell}^{(m)})^2}}.
\end{equation}
For \(M=1\), Equation~\eqref{eq:6.13} reduces to the single-target
second-moment proposal \(\pi_j\propto q_j\sqrt{m_{2j}}\).

\end{proposition}

Second-moment optimality does not control the finite-time range of an
off-policy confidence sequence. For target set \(\mathcal Q\), define
\begin{equation}\label{eq:6.14}
B_{\mathcal Q}(\pi)=\max_{m,j}\frac{q_j^{(m)}}{\pi_j}.
\end{equation}
Imposing \(B_{\mathcal Q}(\pi)\leq C\) is equivalent to lower bounds
\(\pi_j\geq\ell_j(C)=\max_m q_j^{(m)}/C\). Define
\begin{equation}\label{eq:6.15}
C_{\mathcal Q}=\sum_j\max_m q_j^{(m)}.
\end{equation}

\begin{proposition}[Minimum shared importance-weight range]\label{prop:min-range}
For every shared proposal \(\pi\),
\[
B_{\mathcal Q}(\pi)\geq C_{\mathcal Q},
\]
with equality attained by
\[
\pi_j^{\mathrm{range}}=\frac{\max_m q_j^{(m)}}{C_{\mathcal Q}}.
\]
Consequently, \(B_{\mathcal Q}(\pi)\leq C\) is feasible if and only if
\(C\geq C_{\mathcal Q}\). Moreover, \(1\leq C_{\mathcal Q}\leq M\);
\(C_{\mathcal Q}=1\) if and only if all target distributions are identical.
Thus \(C_{\mathcal Q}\) is an intrinsic target-set heterogeneity factor: it is
the smallest worst-case importance weight any one shared proposal can achieve.

\end{proposition}

\begin{corollary}[Range-capped water-filling proposal]\label{cor:waterfill}
For any \(C\geq C_{\mathcal Q}\), there exists a least-favourable
\(\lambda_C^*\in\Delta_M\) such that the minimax second-moment
proposal under the cap \(B_{\mathcal Q}(\pi)\leq C\) has the form
\begin{equation}\label{eq:6.16}
\pi_j^*(C)=\max\left\{\ell_j(C),\;\kappa_C
\sqrt{m_{2j}\sum_m\lambda_{C,m}^*(q_j^{(m)})^2}\right\},
\end{equation}
where \(\kappa_C>0\) is chosen so that \(\sum_j\pi_j^*(C)=1\).
For one target this becomes
\(\pi_j(C)=\max\{q_j/C,\kappa_Cq_j\sqrt{m_{2j}}\}\), a
water-filling interpolation between target sampling and the unconstrained
second-moment proposal. At the left endpoint \(C=C_{\mathcal Q}\), the
proposal is the unique range-minimizing design above; as the cap is relaxed, the
frontier approaches the unconstrained minimax second-moment design.

\end{corollary}

The proof of \cref{prop:minimax-proposal} and \cref{cor:waterfill} is given in Appendix~G.
For the bounded, rescaled PrPl-EB implementation used in the replay,
the cap parameter exposes a finite-sample tradeoff hidden by second-moment
optimization alone: smaller \(C\) reduces the range of the rescaled
importance-weighted stream at the price of a larger worst-case second moment.
This range dependence is not intrinsic to all anytime-valid off-policy
procedures: bound-free betting constructions can remain valid without a known
uniform importance-weight bound (Waudby-Smith et al. 2024a). \cref{prop:min-range}
should therefore be read as an exact geometric statement about the smallest
achievable shared importance-weight range, while the empirical benefit of
range capping in Tables~11--12 is specific to the bounded rescaled PrPl-EB
implementation. Because the proposal is fixed before the replay, no additional
adaptation argument is needed; predictable pilot-based selection of \(C\) can
also be handled by off-policy confidence-sequence methods when the realized
propensities are logged.

\begin{corollary}[Two-target total-variation identity]\label{cor:tv}
For two target distributions \(Q^{(1)}\) and \(Q^{(2)}\),
\[
C_{\{Q^{(1)},Q^{(2)}\}}
=1+\operatorname{TV}\!\left(Q^{(1)},Q^{(2)}\right)
=1+\frac12\sum_j\left|q_j^{(1)}-q_j^{(2)}\right|.
\]
\end{corollary}

Thus the smallest possible worst-case importance-weight inflation required by
one proposal serving two targets is exactly one plus their total-variation
distance. This gives \(C_{\mathcal Q}\) a direct geometric interpretation: two
nearly identical targets admit an almost target-sampling proposal, whereas
well-separated targets necessarily force larger off-policy weights.

\subsection{Multiplicity-free robust directional certification}\label{robust-directional}

The target-set confidence sequences in \cref{prop:targetset-pooled} are designed for simultaneous
reporting of all vertex means, and therefore use a split error budget when built
separately. A weaker but often scientifically central goal is to certify that the
entire target set has the same direction. Because a linear functional achieves its
minimum and maximum over a convex hull at a vertex, robust positivity is the
intersection alternative
\[
H_{1,+}=\{\Delta_{Q^{(m)}}>0\ \text{for every }m\},
\]
against the union null
\[
H_{0,+}=\bigcup_{m=1}^M\{\Delta_{Q^{(m)}}\leq0\}.
\]
This is an intersection--union testing problem (Berger 1982).

\begin{remark}[Intersection--union robust directional certification (Berger 1982)]\label{rem:iut}
For every vertex $m$, let $L_t^{(m)}$ be a one-sided anytime-valid lower
confidence sequence satisfying
\[
\Pr\{L_t^{(m)}\leq\Delta_{Q^{(m)}}\ \text{for every }t\}\geq1-\alpha.
\]
Define
\[
\tau_+=\inf\{t:\min_m L_t^{(m)}>0\}.
\]
Then
\[
\sup_{P\in H_{0,+}}\Pr_P(\tau_+<\infty)\leq\alpha,
\]
with no division of $\alpha$ by $M$. The analogous statement holds for
robust negativity using upper confidence sequences. If a procedure may declare
either robust positivity or robust negativity, running the two directional
families at level $\alpha/2$ controls the probability of any wrong robust-sign
declaration by $\alpha$.

\end{remark}

The result does not provide simultaneous $1-\alpha$ confidence intervals for all
$M$ target means; it provides the weaker decision guarantee that every target in
$\mathcal Q$ has the certified sign. This distinction creates another explicit
point on the guarantee--efficiency frontier: full target-wise reporting may pay
multiplicity, whereas an intersection--union directional decision need not.

\subsection{Gap-aware shared proposals for robust stopping}\label{gap-aware}

Second-moment minimization is target-agnostic with respect to the magnitude of the
effect being certified. For directional stopping, the target gaps matter. Let
\[
g_m=|\Delta_{Q^{(m)}}|>0,
\qquad
V_m(\pi)=S_m(\pi)-\Delta_{Q^{(m)}}^2.
\]
For a root-$n$ target-level confidence sequence with common first-order constant
$\kappa$, the leading stopping scale for target $m$ is proportional to
$V_m(\pi)/g_m^2$. Since
\[
\frac{V_m(\pi)}{g_m^2}
=\frac{S_m(\pi)}{g_m^2}-1,
\]
minimizing the worst first-order stopping scale is equivalent to
\begin{equation}\label{eq:6.17}
\min_{\pi\in\Delta_J}\max_m
\frac{S_m(\pi)}{g_m^2}.
\end{equation}

\begin{proposition}[Gap-aware minimax shared proposal]\label{prop:gap-aware}
Assume $g_m>0$ for every target vertex and $m_{2j}>0$ on every relevant
stratum. There exists a least-favourable vector
$\lambda^\dagger\in\Delta_M$, supported on targets active at the optimum,
such that the solution of Equation~\eqref{eq:6.17} is
\begin{equation}\label{eq:6.18}
\pi_j^\dagger=
\frac{\sqrt{m_{2j}\sum_m\lambda_m^\dagger
(q_j^{(m)})^2/g_m^2}}
{\sum_\ell\sqrt{m_{2\ell}\sum_m\lambda_m^\dagger
(q_\ell^{(m)})^2/g_m^2}}.
\end{equation}
The optimal value has the dual representation
\begin{equation}\label{eq:6.19}
\inf_\pi\max_m\frac{S_m(\pi)}{g_m^2}
=
\max_{\lambda\in\Delta_M}
\left[
\sum_j\sqrt{m_{2j}\sum_m\lambda_m
(q_j^{(m)})^2/g_m^2}
\right]^2.
\end{equation}
For a range cap $B_{\mathcal Q}(\pi)\leq C$, the same KKT argument gives
the water-filling form in \cref{cor:waterfill} with the coefficients
$(q_j^{(m)})^2m_{2j}$ replaced by
$(q_j^{(m)})^2m_{2j}/g_m^2$.

\end{proposition}

Thus the second-moment proposal in \cref{prop:minimax-proposal} and the gap-aware proposal
solve different design problems: the former protects raw target-wise variance,
whereas the latter protects the target that is hardest to certify relative to its
gap. This criterion is closer to the signal-to-noise allocation logic in
Demonstration Experiments (Imbens et al. 2026), but here the competing objects
are externally weighted target mixtures rather than arms.

\subsection{Adaptive tracking of the gap-aware proposal}\label{gap-aware-tracking}

The gaps and second moments in \cref{prop:gap-aware} are generally unknown. They can
be learned while preserving exploration. Let $\widehat\delta_{j,t}$ and
$\widehat m_{2j,t}$ be the within-stratum sample mean and second-moment estimates,
and let
\[
\widehat g_{m,t}=\max\left\{\left|\sum_jq_j^{(m)}
\widehat\delta_{j,t}\right|,\gamma_t\right\},
\]
where $\gamma_t\downarrow0$. Let $\widehat\pi_t^\dagger$ solve the plug-in
version of Equation~\eqref{eq:6.17}, and sample at the next time from
\[
\widetilde\pi_{t+1}=(1-\varepsilon_t)\widehat\pi_t^\dagger
+\varepsilon_t\nu,
\]
where $\nu_j>0$, $\varepsilon_t\downarrow0$, and
$\sum_t\varepsilon_t=\infty$.

\begin{theorem}[Adaptive gap-aware proposal tracking]\label{thm:gap-tracking}
Suppose outcomes within each stratum are i.i.d. and bounded, every target gap
$g_m$ is nonzero, all target-relevant second moments are positive, and the
population minimax problem in Equation~\eqref{eq:6.17} has a unique solution
$\pi^\dagger$. Then
\begin{equation}\label{eq:6.20}
\widehat\pi_t^\dagger\to\pi^\dagger,
\qquad
\widetilde\pi_t\to\pi^\dagger,
\qquad
\frac{N_j(t)}{t}\to\pi_j^\dagger
\quad\text{a.s. for every }j.
\end{equation}
\end{theorem}

The proof combines infinite exploration, strong consistency of local moments,
continuity of the unique minimax solution, and the martingale strong law. As in
\cref{thm:tracking}, the exploration floor guarantees learning while vanishing as a
fraction of the total budget.

\subsection{Predictable centering and efficient pooled variance}\label{predictable-centering}

The raw importance-weighted process in Equation~\eqref{eq:6.10} is unbiased, but its
second moment contains both within-stratum variance and between-stratum mean
heterogeneity. The augmented score below is the stratum-level specialization
of the doubly robust anytime-valid off-policy construction developed by
Waudby-Smith et al. (2024a): because the logging propensities are design-known,
a predictable outcome model can be used as a control variate while retaining
exact conditional unbiasedness. In this specialization, the control variate
removes the between-stratum mean contribution as the local mean estimates
stabilize. Let
\(\widehat\delta_{j,t-1}\) be any estimate based only on observations available
before time \(t\), and define for target vertex \(m\)
\begin{equation}\label{eq:6.21}
Y_t^{(m)}=
\sum_{j=1}^J q_j^{(m)}\widehat\delta_{j,t-1}
+\frac{q_{A_t}^{(m)}}{\pi_t(A_t)}
\left(D_t-\widehat\delta_{A_t,t-1}\right).
\end{equation}

\begin{proposition}[Specialized augmented off-policy score and variance decomposition]\label{prop:centering}
For every predictable proposal \(\pi_t\) satisfying positivity,
\[
\mathbb E\!\left(Y_t^{(m)}\mid\mathcal H_{t-1}\right)
=\Delta_{Q^{(m)}}.
\]
Writing \(r_{m,t-1}=\sum_jq_j^{(m)}
(\delta_j-\widehat\delta_{j,t-1})\), its conditional variance is
\begin{equation}\label{eq:6.22}
\operatorname{Var}\!\left(Y_t^{(m)}\mid\mathcal H_{t-1}\right)
=
\sum_{j=1}^J\frac{(q_j^{(m)})^2}{\pi_t(j)}
\left\{\sigma_j^2+
(\delta_j-\widehat\delta_{j,t-1})^2\right\}
-r_{m,t-1}^2.
\end{equation}
If \(\widehat\delta_{j,t}\to\delta_j\) and
\(\pi_t(j)\to\pi_j>0\), then the conditional variance converges almost surely to
\begin{equation}\label{eq:6.23}
V_m^{\mathrm{ctr}}(\pi)=
\sum_{j=1}^J\frac{(q_j^{(m)})^2\sigma_j^2}{\pi_j}.
\end{equation}

For one target, minimizing Equation~\eqref{eq:6.23} over the simplex gives
\begin{equation}\label{eq:6.24}
\pi_j^{\mathrm{ctr},*}
=\frac{q_j\sigma_j}{\sum_\ell q_\ell\sigma_\ell}.
\end{equation}
\end{proposition}

Thus the Neyman structure, which is generally not optimal for the additive
local-CS width objective, re-emerges as the asymptotically variance-efficient
allocation for predictably centered pooled target inference. This is another
sense in which the optimal allocation is architecture-specific. For several
targets, replacing \(m_{2j}\) by \(\sigma_j^2\) in \cref{prop:gap-aware} gives the
centered gap-aware shared proposal.

\subsection{Gaussian information limit for robust target-set certification}\label{information-limit}

The preceding design criteria arise from confidence-sequence width. A stronger
benchmark asks what any sequential procedure must pay, regardless of the
particular confidence-sequence construction. Consider independent Gaussian
strata
\[
D_{j,r}\sim N(\delta_j,\sigma_j^2),\qquad \sigma_j^2>0
\]
with known variances, and suppose the true parameter satisfies
\(g_m=(Q^{(m)})^\top\delta>0\) for every target vertex. Let a sequential
procedure choose strata predictably, stop at \(\tau\), and make a robust-positive
declaration. Call it \((\alpha,\beta)\)-valid at \(\delta\), with
\(0<\alpha<1-\beta\), if the declaration probability is at least
\(1-\beta\) under \(\delta\) and at most \(\alpha\) under every parameter in
the robust null
\[
\mathcal H_{0,+}=\bigcup_{m=1}^M
\{\delta':(Q^{(m)})^\top\delta'\leq0\}.
\]
For \(w\in\Delta_J\), define
\begin{equation}\label{eq:6.25}
V_m(w)=\sum_{j=1}^J\frac{(q_j^{(m)})^2\sigma_j^2}{w_j},
\qquad
\Gamma^*(\delta,\mathcal Q)
=\max_{w\in\Delta_J}\min_m\frac{g_m^2}{2V_m(w)}.
\end{equation}

\begin{theorem}[Information-theoretic lower bound]\label{thm:lower-bound}
Assume \(\mathbb E_\delta\tau<\infty\). Every
\((\alpha,\beta)\)-valid robust-positive procedure satisfies
\begin{equation}\label{eq:6.26}
\mathbb E_\delta\tau
\geq
\frac{\operatorname{kl}(1-\beta,\alpha)}
{\Gamma^*(\delta,\mathcal Q)},
\end{equation}
where \(\operatorname{kl}(p,q)\) is the binary relative entropy. In particular,
for procedures that declare robust positivity almost surely under the
alternative,
\begin{equation}\label{eq:6.27}
\mathbb E_\delta\tau
\geq
\frac{\log(1/\alpha)}{\Gamma^*(\delta,\mathcal Q)}.
\end{equation}
\end{theorem}

The bound adapts the bandit change-of-measure inequality of
Kaufmann, Capp\'e, and Garivier (2016) to the target-set null geometry. For a
fixed allocation \(w\), the closest Gaussian parameter satisfying
\((Q^{(m)})^\top\delta'\leq0\) has information distance
\[
\inf_{(Q^{(m)})^\top\delta'\leq0}
\sum_j w_j\,
\mathrm{KL}\!\left(
N(\delta_j,\sigma_j^2),N(\delta'_j,\sigma_j^2)
\right)
=
\frac{g_m^2}{2V_m(w)}.
\]

\begin{corollary}[Information-optimal shared allocation]\label{cor:info-optimal}
Let
\[
D^*=\min_{w\in\Delta_J}\max_m\frac{V_m(w)}{g_m^2}.
\]
Then \(\Gamma^*=1/(2D^*)\). There exists
\(\lambda^*\in\Delta_M\), supported on active worst-case targets, such that
\begin{equation}\label{eq:6.28}
w_j^*=
\frac{\sigma_j
\sqrt{\sum_m\lambda_m^*(q_j^{(m)})^2/g_m^2}}
{\sum_\ell\sigma_\ell
\sqrt{\sum_m\lambda_m^*(q_\ell^{(m)})^2/g_m^2}},
\end{equation}
and
\begin{equation}\label{eq:6.29}
D^*=
\max_{\lambda\in\Delta_M}
\left[
\sum_j\sigma_j
\sqrt{\sum_m\lambda_m(q_j^{(m)})^2/g_m^2}
\right]^2.
\end{equation}
For a single target,
\begin{equation}\label{eq:6.30}
w_j^*=\frac{q_j\sigma_j}{\sum_\ell q_\ell\sigma_\ell},
\qquad
\frac{1}{\Gamma^*}
=\frac{2(\sum_jq_j\sigma_j)^2}{\Delta_Q^2}.
\end{equation}
\end{corollary}

Hence the same centered gap-aware design that minimizes the first-order pooled
variance-to-gap criterion also maximizes the information rate in this Gaussian
benchmark.

\begin{proposition}[Sufficient conditions for first-order attainment]\label{prop:attainment}
Let \(\alpha\downarrow0\), and suppose an anytime-valid robust-positive procedure
satisfies both: (i) \(N_j(t)/t\to w_j^*\) in probability uniformly on the
\(t\asymp\log(1/\alpha)\) stopping scale; and (ii) for every target vertex its
one-sided lower confidence boundary has the first-order expansion
\[
L_{m,t}=\widehat g_{m,t}-
\sqrt{\frac{2V_m(w^*)}{t}
\{\log(1/\alpha)+o_{\mathbb P}(\log(1/\alpha))\}},
\]
with \(\widehat g_{m,t}\to g_m\) uniformly on that scale. If the procedure stops
when \(\min_mL_{m,t}>0\), then
\begin{equation}\label{eq:6.31}
\frac{\tau_\alpha}{\log(1/\alpha)}
\xrightarrow{\mathbb P}
\frac{1}{\Gamma^*(\delta,\mathcal Q)}.
\end{equation}
If, in addition, the normalized stopping times
\(\{\tau_\alpha/\log(1/\alpha)\}\) are uniformly integrable, then
\[
\frac{\mathbb E_\delta\tau_\alpha}{\log(1/\alpha)}
\longrightarrow
\frac{1}{\Gamma^*(\delta,\mathcal Q)},
\]
so the expected-sample-size lower-bound constant is attained as well. \end{proposition}

Thus these conditions are sufficient for first-order attainment of the
Gaussian lower-bound constant. This proposition deliberately does not claim
that the bounded PrPl-EB implementation used in the replay satisfies the sharp
boundary expansion; establishing that fact for a concrete finite-sample-valid
bounded-data procedure is a separate problem. This proposition separates the information limit from the
finite-sample choice of confidence-sequence machinery; Gaussian mixture
boundaries and asymptotic confidence-sequence constructions provide examples of
families with the required first-order form under appropriate regularity
conditions (Howard et al. 2021; Waudby-Smith et al. 2024b).

\subsection{Exact ternary-family information benchmark}\label{ternary-information}

The Gaussian calculation in Section~6.13 is useful because it yields a closed-form
allocation, but the PromptEval replay itself uses bounded ternary differences.  We
therefore also compute the change-of-measure characteristic information in the
exact reconstructed family on \(\{-1,0,1\}\).  Let \(P_j\) denote the true
ternary law in stratum \(j\), let \(x\in\{-1,0,1\}\), and for a target vertex
\(m\) define
\begin{equation}\label{eq:6.32}
I_m^{\mathrm{ter}}(w)=
\inf_{\{R_j\}:\;\sum_j q_j^{(m)}E_{R_j}D\leq0}
\sum_j w_j\,\mathrm{KL}(P_j\Vert R_j),
\qquad
\Gamma_{\mathrm{ter}}^*=\max_{w\in\Delta_J}\min_m I_m^{\mathrm{ter}}(w).
\end{equation}

\begin{proposition}[Ternary-family I-projection and lower bound]\label{prop:ternary}
For fixed \(w_j>0\) and target vertex \(m\) with positive true gap, the minimizer
in Equation~\eqref{eq:6.32} lies on the boundary
\(\sum_jq_j^{(m)}E_{R_j}D=0\).  There is a multiplier \(\lambda\geq0\) and,
for each stratum, a normalizing constant \(c_j\) such that
\begin{equation}\label{eq:6.33}
R_j(x)=
\frac{P_j(x)}{c_j+(\lambda q_j^{(m)}/w_j)x},
\qquad x\in\{-1,0,1\},
\end{equation}
where \(c_j\) is the unique value making \(R_j\) sum to one and \(\lambda\)
is chosen to satisfy the target-mean boundary.  Consequently, the standard
sequential change-of-measure inequality gives
\begin{equation}\label{eq:6.34}
E_P\tau\geq
\frac{\operatorname{kl}(1-\beta,\alpha)}{\Gamma_{\mathrm{ter}}^*}
\end{equation}
for every robust-positive procedure that is \((\alpha,\beta)\)-valid over the
ternary model. In particular, the bound applies to every procedure that is
valid for all stratum laws supported in \([-1,1]\), because such a procedure is
valid over the ternary sub-model.

\end{proposition}

\begin{lemma}[Equivalence with the bounded nonparametric information]\label{lem:klinf}
Suppose each \(P_j\) is supported in \(\{-1,0,1\}\). Then the infimum in
Equation~\eqref{eq:6.32} is unchanged if each \(R_j\) ranges over all
probability laws on \([-1,1]\) rather than over laws on \(\{-1,0,1\}\).
Consequently \(\Gamma^*_{\mathrm{ter}}\) equals the stratified
\(\mathrm{KL}_{\inf}\) characteristic information for
\([-1,1]\)-bounded outcomes, and Equation~\eqref{eq:6.34} is the
nonparametric bounded-data lower bound for the replay population.
\end{lemma}

The proof is the reverse-KL KKT projection for a linear mean constraint followed
by the same adaptive change-of-measure argument used in \cref{thm:lower-bound}.  The
accompanying script solves the outer max--min problem numerically in softmax
coordinates, using the envelope gradient supplied by the I-projection.

\section{Simultaneous best-system
identification}\label{simultaneous-best-system-identification}

Let \(K\) systems be indexed by \(k = 1,\ldots,K\), and let

\[\theta_{k,Q} = \sum_{j}^{}q_{j}\mu_{k,j}\]

be the target performance of system \(k\). For each unordered pair
\(h = (a,b)\), define

\[\Delta_{ab,Q} = \theta_{a,Q} - \theta_{b,Q}.\]

There are

\[H = \binom{K}{2}\]

pairs. Construct local pair-by-stratum confidence sequences at error
level

\[\alpha_{0} = \frac{\alpha}{HJ}.\]

By \cref{thm:target-interval} and a union bound over pairs, all pairwise target
sequences cover simultaneously for all times with probability at least
\(1 - \alpha\).

At time \(t\), identify system \(w\) as uniquely best if

\begin{equation}\label{eq:7.1}
L_{wk,Q,t} > 0\quad\quad\text{for every }k \neq w,
\end{equation}

where \(L_{wk,Q,t}\) is the lower confidence bound for
\(\theta_{w,Q} - \theta_{k,Q}\), using the sign-reversed upper bound
when the stored unordered pair is \((k,w)\).

\begin{corollary}[Familywise anytime-valid best-system identification]\label{cor:best-system} Under the simultaneous construction above,

\begin{equation}\label{eq:7.2}
\mathbb{P}\left( \exists t,\exists w:\mspace{6mu} w\text{ is identified as uniquely best at }t\text{ and }\min_{k \neq w}\Delta_{wk,Q} \leq 0 \right) \leq \alpha.
\end{equation}

\end{corollary}

Equation~\eqref{eq:7.2} is an ``ever'' statement: it controls the probability
that the procedure makes an incorrect identification at any monitored
time, not merely at a fixed analysis time.

For an uncertain target set \(\mathcal{Q}\), replace each pairwise lower
bound by

\[L_{ab,\mathcal{Q},t} = \min_{m}\sum_{j}^{}q_{j}^{(m)}L_{ab,j,t}.\]

System \(w\) is \emph{robustly} identified as best when
\(L_{wk,\mathcal{Q},t} > 0\) for every \(k \neq w\). \cref{prop:uniform-targetset}
implies the same familywise error bound using the original
\(\alpha/(HJ)\) local split; no additional division by \(M\) is
necessary because all target scenarios are deterministic linear
combinations on the same simultaneous pair--stratum coverage event.

The Bonferroni construction is intentionally simple and can be
conservative when \(K\) is large. Fixed-confidence best-arm
identification provides a mature literature on sharper allocation and
stopping rules, including information-theoretic lower bounds and
Track-and-Stop (Kaufmann, Cappé, and Garivier 2016; Garivier and
Kaufmann 2016). Our setting differs in that the sampling decision may be
over evaluation conditions and the target is an externally weighted
mixture \(Q\), while a single selected condition may yield paired
information about several systems. Developing asymptotically optimal
allocation under this combined structure is an important extension
beyond the present conservative identification rule. Earlier
pure-exploration work likewise shows that allocation and stopping can be
coupled more tightly than a simple union-bound construction permits
(Audibert, Bubeck, and Munos 2010; Jamieson et al. 2014).

\section{Simulation studies}\label{simulation-studies}

The original design experiments are reproduced by the accompanying
base-R script \path{reproduce_target_aware_ai_eval.R}. The
variance-adaptive, optimal-allocation-tracking, and robust-target
extensions are reproduced by
\path{reproduce_variance_adaptive_upgrade_fairbatch.R}. The
variance-adaptive script was run in R 4.3.3 with fixed seeds, and
Tables~4--6 report those shipped outputs. The pooled architecture is
reproduced by \path{reproduce_direct_pooled_cs.R}. The canonical
pooled script uses five pilot observations, 5,000 Monte Carlo
trajectories, and fixed seed 20260908. We report batch-100 monitoring
for the PromptEval architecture comparison so that its monitoring
granularity matches Table~9, and we report batch-10 results as a
sensitivity analysis.

\subsection{Data-generating process}\label{data-generating-process}

We use \(J = 4\) strata with target weights

\[Q = (0.55,0.20,0.15,0.10)\]

and pairwise stratum gaps

\[\delta = (0.08,0.04,0.12,0.03),\]

so

\[\Delta_{Q} = 0.073.\]

For the bounded experiments, \(D_{j} \in \{ - 1,0,1\}\) with

\[\mathbb{P}\left( D_{j} = 1 \right) = \frac{s_{j} + \delta_{j}}{2},\quad\mathbb{P}\left( D_{j} = - 1 \right) = \frac{s_{j} - \delta_{j}}{2},\quad\mathbb{P}\left( D_{j} = 0 \right) = 1 - s_{j},\]

where

\[s = (0.15,0.80,0.50,0.95).\]

Thus

\[\sigma_{j}^{2} = s_{j} - \delta_{j}^{2} = (0.1436,0.7984,0.4856,0.9491).\]

The design intentionally separates target prevalence from variability:
the largest target stratum is relatively stable, while several
lower-weight strata are much noisier. This is not intended to mimic a
specific benchmark. It creates a transparent setting in which
variance-optimal and width-optimal objectives disagree.

\subsection{Fixed-budget objective
mismatch}\label{fixed-budget-objective-mismatch}

Table~1 uses a total budget of 400 paired evaluation units and 20,000
Monte Carlo replications. ``Range-width'' uses allocation weights
\(q_{j}^{2/3}\) and ``variance-width'' uses
\(\left( q_{j}\sigma_{j} \right)^{2/3}\). The width columns report the
leading proxies

\[W_{range} = \sum_{j}^{}\frac{q_{j}}{\sqrt{n_{j}}},\quad\quad W_{var} = \sum_{j}^{}\frac{q_{j}\sigma_{j}}{\sqrt{n_{j}}}.\]

\textbf{Table 1. Fixed-budget estimation and width objectives (B=400)}

\begin{longtable}[]{@{}
  >{\raggedright\arraybackslash}p{(\columnwidth - 8\tabcolsep) * \real{0.2000}}
  >{\raggedright\arraybackslash}p{(\columnwidth - 8\tabcolsep) * \real{0.2000}}
  >{\raggedright\arraybackslash}p{(\columnwidth - 8\tabcolsep) * \real{0.2000}}
  >{\raggedright\arraybackslash}p{(\columnwidth - 8\tabcolsep) * \real{0.2000}}
  >{\raggedright\arraybackslash}p{(\columnwidth - 8\tabcolsep) * \real{0.2000}}@{}}
\toprule\noalign{}
\begin{minipage}[b]{\linewidth}\raggedright
\textbf{Design}
\end{minipage} & \begin{minipage}[b]{\linewidth}\raggedright
\textbf{Analytic variance}
\end{minipage} & \begin{minipage}[b]{\linewidth}\raggedright
\textbf{MC variance}
\end{minipage} & \begin{minipage}[b]{\linewidth}\raggedright
\textbf{Wrange}
\end{minipage} & \begin{minipage}[b]{\linewidth}\raggedright
\textbf{Wvar}
\end{minipage} \\
\midrule\noalign{}
\endhead
\bottomrule\noalign{}
\endlastfoot
Uniform & 0.000958 & 0.000957 & 0.100000 & 0.058908 \\
Q-proportional & 0.001016 & 0.001020 & 0.094618 & 0.062930 \\
Neyman estimation & \textbf{0.000868} & \textbf{0.000856} & 0.094448 &
0.058133 \\
Range-width identification & 0.000911 & 0.000923 & \textbf{0.092872} &
0.059433 \\
Variance-width identification & 0.000877 & 0.000855 & 0.095430 &
\textbf{0.057875} \\
\end{longtable}

The reversal is the key result. \(Q\)-proportional sampling improves the
range-only width relative to uniform sampling but has \emph{higher}
point-estimator variance in this DGP because it undersamples
high-variance strata. Neyman allocation is best for estimation, while
\(q^{2/3}\) is best for the range-width proxy. Therefore statements such
as ``sample proportional to deployment frequency'' are incomplete
without specifying the statistical objective and heterogeneity
structure.

\subsection{Anytime-valid pairwise
identification}\label{anytime-valid-pairwise-identification}

We next use the bounded normal-mixture reference boundary in
Equation~\eqref{eq:6.1}, with \(\alpha = .05\), local level \(\alpha/J\),
\(\rho = 200\), 20 initial observations per stratum, and batches of 100.
Table~2 reports 2,000 Monte Carlo runs. The fixed allocation shares
determine how each new batch is distributed.

\textbf{Table 2. Paired evaluation units to bounded range-only
anytime-valid pairwise identification}

\begin{longtable}[]{@{}
  >{\raggedright\arraybackslash}p{(\columnwidth - 8\tabcolsep) * \real{0.2000}}
  >{\raggedright\arraybackslash}p{(\columnwidth - 8\tabcolsep) * \real{0.2000}}
  >{\raggedright\arraybackslash}p{(\columnwidth - 8\tabcolsep) * \real{0.2000}}
  >{\raggedright\arraybackslash}p{(\columnwidth - 8\tabcolsep) * \real{0.2000}}
  >{\raggedright\arraybackslash}p{(\columnwidth - 8\tabcolsep) * \real{0.2000}}@{}}
\toprule\noalign{}
\begin{minipage}[b]{\linewidth}\raggedright
\textbf{Design}
\end{minipage} & \begin{minipage}[b]{\linewidth}\raggedright
\textbf{Median}
\end{minipage} & \begin{minipage}[b]{\linewidth}\raggedright
\textbf{25th pct.}
\end{minipage} & \begin{minipage}[b]{\linewidth}\raggedright
\textbf{75th pct.}
\end{minipage} & \begin{minipage}[b]{\linewidth}\raggedright
\textbf{Mean}
\end{minipage} \\
\midrule\noalign{}
\endhead
\bottomrule\noalign{}
\endlastfoot
Uniform & 9,180 & 8,180 & 10,380 & 9,274.4 \\
Q-proportional & 8,280 & 7,380 & 9,380 & 8,382.4 \\
Neyman estimation & 8,280 & 7,280 & 9,280 & 8,315.8 \\
Range-width identification & \textbf{7,980} & 7,080 & 8,980 &
\textbf{8,086.2} \\
Variance-width identification & 8,280 & 7,380 & 9,280 & 8,376.4 \\
\end{longtable}

For a pairwise comparison, each evaluation unit corresponds to two model
calls under the simple accounting used here; thus the median 7,980
evaluation units for the range-width design correspond to 15,960 model
calls. The range-width allocation is best under the range-only stopping
boundary. This is precisely the objective mismatch predicted by
\cref{prop:boundary-rate}. The result also explains why a variance-estimation
adaptive rule can appear successful under a range-only stopping
criterion without actually optimizing the criterion: its performance may
be driven mainly by the allocation shares it happens to approach.

\subsection{Boundary choice dominates allocation
choice}\label{boundary-choice-dominates-allocation-choice}

The range-only sequence treats all strata according to the common bound
\(\lbrack - 1,1\rbrack\), even though their variances differ by more
than a factor of six. To calibrate the potential value of variance
adaptation, we repeat the sequential experiment under a variance-known
Gaussian model with the same \(\delta_{j}\) and \(\sigma_{j}^{2}\),
using the known variance in a normal-mixture boundary. This is a
\emph{variance-known benchmark}, not a claim that the Gaussian boundary
is a valid finite-sample replacement for the bounded DGP. Practical
bounded-data implementations should use empirical-Bernstein or betting
confidence sequences (Waudby-Smith and Ramdas 2024).

\textbf{Table 3. Variance-known Gaussian benchmark (5,000 replications)}

\begin{longtable}[]{@{}
  >{\raggedright\arraybackslash}p{(\columnwidth - 8\tabcolsep) * \real{0.2000}}
  >{\raggedright\arraybackslash}p{(\columnwidth - 8\tabcolsep) * \real{0.2000}}
  >{\raggedright\arraybackslash}p{(\columnwidth - 8\tabcolsep) * \real{0.2000}}
  >{\raggedright\arraybackslash}p{(\columnwidth - 8\tabcolsep) * \real{0.2000}}
  >{\raggedright\arraybackslash}p{(\columnwidth - 8\tabcolsep) * \real{0.2000}}@{}}
\toprule\noalign{}
\begin{minipage}[b]{\linewidth}\raggedright
\textbf{Design}
\end{minipage} & \begin{minipage}[b]{\linewidth}\raggedright
\textbf{Median}
\end{minipage} & \begin{minipage}[b]{\linewidth}\raggedright
\textbf{25th pct.}
\end{minipage} & \begin{minipage}[b]{\linewidth}\raggedright
\textbf{75th pct.}
\end{minipage} & \begin{minipage}[b]{\linewidth}\raggedright
\textbf{Mean}
\end{minipage} \\
\midrule\noalign{}
\endhead
\bottomrule\noalign{}
\endlastfoot
Uniform & 3,280 & 2,580 & 3,980 & 3,330.6 \\
Q-proportional & 3,580 & 2,955 & 4,380 & 3,713.1 \\
Neyman estimation & \textbf{3,180} & 2,580 & 3,780 & 3,234.9 \\
Range-width identification & 3,280 & 2,680 & 3,980 & 3,342.8 \\
Variance-width identification & \textbf{3,180} & 2,580 & 3,780 &
\textbf{3,221.0} \\
\end{longtable}

Changing allocation within the range-only procedure saved roughly 1,200
median evaluation units (9,180 to 7,980). Moving from the conservative
range-only boundary to the variance-known Gaussian benchmark changes the
scale much more dramatically, to about 3,100--3,300 median evaluation
units for the better allocations. The practical lesson is that
allocation and confidence-sequence construction must be optimized
jointly; an elegant sampling policy cannot compensate for a severely
conservative boundary.

\subsubsection{Reproducibility note}\label{reproducibility-note}

Tables~1--3 report the canonical base-R run with seed 20260831. All
reported Monte Carlo summaries in these tables are generated from that
fixed-seed R workflow.

\subsection{Finite-sample variance-adaptive
identification}\label{finite-sample-variance-adaptive-identification}

We next replace the variance-known Gaussian benchmark by the
finite-sample PrPl-EB confidence sequence in Equation~\eqref{eq:6.2}. Outcomes
remain the bounded ternary differences defined in the data-generating
process above. We use local level \(\alpha/J\), \(c = 1/2\), 20 pilot
observations per stratum, 2,000 Monte Carlo replications, and a common
monitoring/allocation batch of 40 for \emph{all} fixed and adaptive
policies. The adaptive policy uses Equation~\eqref{eq:6.4}. Table~4 reports
evaluation units to pairwise identification.

\textbf{Table 4. Finite-sample variance-adaptive PrPl-EB pairwise
identification (2,000 replications)}

\begin{longtable}[]{@{}
  >{\raggedright\arraybackslash}p{(\columnwidth - 10\tabcolsep) * \real{0.1667}}
  >{\raggedright\arraybackslash}p{(\columnwidth - 10\tabcolsep) * \real{0.1667}}
  >{\raggedright\arraybackslash}p{(\columnwidth - 10\tabcolsep) * \real{0.1667}}
  >{\raggedright\arraybackslash}p{(\columnwidth - 10\tabcolsep) * \real{0.1667}}
  >{\raggedright\arraybackslash}p{(\columnwidth - 10\tabcolsep) * \real{0.1667}}
  >{\raggedright\arraybackslash}p{(\columnwidth - 10\tabcolsep) * \real{0.1667}}@{}}
\toprule\noalign{}
\begin{minipage}[b]{\linewidth}\raggedright
\textbf{Design}
\end{minipage} & \begin{minipage}[b]{\linewidth}\raggedright
\textbf{Median}
\end{minipage} & \begin{minipage}[b]{\linewidth}\raggedright
\textbf{25th pct.}
\end{minipage} & \begin{minipage}[b]{\linewidth}\raggedright
\textbf{75th pct.}
\end{minipage} & \begin{minipage}[b]{\linewidth}\raggedright
\textbf{Mean}
\end{minipage} & \begin{minipage}[b]{\linewidth}\raggedright
\textbf{Unresolved}
\end{minipage} \\
\midrule\noalign{}
\endhead
\bottomrule\noalign{}
\endlastfoot
Uniform & 4,680 & 3,800 & 5,800 & 4,845.64 & 0.0000 \\
Q-proportional & 5,280 & 4,280 & 6,520 & 5,469.04 & 0.0025 \\
Neyman estimation & 4,560 & 3,640 & 5,600 & 4,715.10 & 0.0005 \\
Range-width q\^{}(2/3) & 4,720 & 3,760 & 5,880 & 4,912.96 & 0.0000 \\
First-order variance-optimal $(q\sigma)^{2/3}$ & 4,560 & 3,680 & 5,600 &
4,704.52 & 0.0005 \\
Adaptive PrPl-EB & \textbf{4,520} & \textbf{3,560} & 5,610 &
\textbf{4,684.00} & 0.0000 \\
\end{longtable}

The fully valid variance-adaptive construction substantially reduces the
median evaluation burden relative to the transparent range-only
reference procedure (7,980 versus 4,520 for the best design in each
table), while preserving anytime validity. The adaptive policy operates
at essentially the same scale as the first-order variance-optimal
allocation without knowing the stratum variances in advance. The point
is not that 4,520 is universal---it depends on the DGP, boundary and
tuning---but that variance adaptation is operationally available with
finite-sample validity rather than only as a variance-known Gaussian
benchmark.

The pooled script also evaluates this synthetic DGP. With five pooled
pilot observations, batch-10 monitoring, and 5,000 trajectories, direct
target sampling stops at a median of 965 units (IQR 535--1,565), whereas
the second-moment proposal in Equation~\eqref{eq:3.5} stops at 765 (IQR
475--1,125) with importance-weight bound \(B(\pi) = 1.54\). Relative to
the best stratum-resolved median in Table~4, these values correspond to
reductions of 78.7\% and 83.1\%, respectively, although the monitoring
batches differ. Under batch-100 monitoring the corresponding medians are
1,005 and 805. Thus the second-moment proposal outperforms direct target
sampling in this four-stratum DGP, in contrast to the PromptEval replay
below. The reversal is useful: range inflation from importance weighting
is a finite-sample tradeoff, not a general reason for the second-moment
proposal to lose.

\subsection{Optimal-allocation tracking
diagnostics}\label{optimal-allocation-tracking-diagnostics}

\cref{thm:tracking} is asymptotic, so we separately examine allocation shares at
fixed horizons rather than at stopping times. The optimal shares in this
DGP are

\[w^{opt} = (0.3189,0.2878,0.2013,0.1921).\]

With 1,000 replications, four pilot observations per stratum,
\(\nu = Q\), and \(\varepsilon_{t} = \min\{ 1,2t^{- 0.6}\}\), Table~5
shows convergence toward these shares. The mean \(\ell_{1}\) distance is
computed replication-by-replication before averaging.

\textbf{Table 5. Adaptive optimal-allocation tracking at fixed horizons
(1,000 replications)}

\begin{longtable}[]{@{}
  >{\raggedright\arraybackslash}p{(\columnwidth - 10\tabcolsep) * \real{0.1667}}
  >{\raggedright\arraybackslash}p{(\columnwidth - 10\tabcolsep) * \real{0.1667}}
  >{\raggedright\arraybackslash}p{(\columnwidth - 10\tabcolsep) * \real{0.1667}}
  >{\raggedright\arraybackslash}p{(\columnwidth - 10\tabcolsep) * \real{0.1667}}
  >{\raggedright\arraybackslash}p{(\columnwidth - 10\tabcolsep) * \real{0.1667}}
  >{\raggedright\arraybackslash}p{(\columnwidth - 10\tabcolsep) * \real{0.1667}}@{}}
\toprule\noalign{}
\begin{minipage}[b]{\linewidth}\raggedright
\textbf{Horizon}
\end{minipage} & \begin{minipage}[b]{\linewidth}\raggedright
\textbf{$\bar{w}_1$}
\end{minipage} & \begin{minipage}[b]{\linewidth}\raggedright
\textbf{$\bar{w}_2$}
\end{minipage} & \begin{minipage}[b]{\linewidth}\raggedright
\textbf{$\bar{w}_3$}
\end{minipage} & \begin{minipage}[b]{\linewidth}\raggedright
\textbf{$\bar{w}_4$}
\end{minipage} & \begin{minipage}[b]{\linewidth}\raggedright
\textbf{Mean $\ell_1$ error}
\end{minipage} \\
\midrule\noalign{}
\endhead
\bottomrule\noalign{}
\endlastfoot
500 & 0.3196 & 0.2875 & 0.2029 & 0.1899 & 0.0978 \\
2,000 & 0.3247 & 0.2863 & 0.2001 & 0.1889 & 0.0427 \\
10,000 & 0.3223 & 0.2866 & 0.2006 & 0.1906 & 0.0179 \\
\end{longtable}

The diagnostic is deliberately separated from the stopping experiment
because shares observed at a data-dependent stopping time are
selection-biased toward easier Monte Carlo paths and are not a clean
test of \cref{thm:tracking}.

\subsection{Robust allocation under deployment
uncertainty}\label{robust-allocation-under-deployment-uncertainty}

To illustrate \cref{prop:least-favourable}, retain the same stratum variances but
consider three plausible deployment targets,

\[Q^{(1)} = (.55,.20,.15,.10),\quad Q^{(2)} = (.10,.55,.20,.15),\quad Q^{(3)} = (.15,.10,.55,.20).\]

For a root-\(n\) variance-width objective with equal costs and
\(B = 400\), Table~6 compares designs optimized for each single target
with the minimax design over their convex hull. ``Worst width'' is the
largest \(\sum_{j}^{}q_{j}^{(m)}\sigma_{j}/\sqrt{n_{j}}\) across the
three targets.

\textbf{Table 6. Target-specific and robust variance-width allocation
under Q = conv\{Q\^{}(1), Q\^{}(2), Q\^{}(3)\}}

\begin{longtable}[]{@{}
  >{\raggedright\arraybackslash}p{(\columnwidth - 10\tabcolsep) * \real{0.1667}}
  >{\raggedright\arraybackslash}p{(\columnwidth - 10\tabcolsep) * \real{0.1667}}
  >{\raggedright\arraybackslash}p{(\columnwidth - 10\tabcolsep) * \real{0.1667}}
  >{\raggedright\arraybackslash}p{(\columnwidth - 10\tabcolsep) * \real{0.1667}}
  >{\raggedright\arraybackslash}p{(\columnwidth - 10\tabcolsep) * \real{0.1667}}
  >{\raggedright\arraybackslash}p{(\columnwidth - 10\tabcolsep) * \real{0.1667}}@{}}
\toprule\noalign{}
\begin{minipage}[b]{\linewidth}\raggedright
\textbf{Design}
\end{minipage} & \begin{minipage}[b]{\linewidth}\raggedright
\textbf{w1}
\end{minipage} & \begin{minipage}[b]{\linewidth}\raggedright
\textbf{w2}
\end{minipage} & \begin{minipage}[b]{\linewidth}\raggedright
\textbf{w3}
\end{minipage} & \begin{minipage}[b]{\linewidth}\raggedright
\textbf{w4}
\end{minipage} & \begin{minipage}[b]{\linewidth}\raggedright
\textbf{Worst width}
\end{minipage} \\
\midrule\noalign{}
\endhead
\bottomrule\noalign{}
\endlastfoot
Optimal for Q\^{}(1) & 0.3189 & 0.2878 & 0.2013 & 0.1921 & 0.08136 \\
Optimal for Q\^{}(2) & 0.0880 & 0.4858 & 0.2097 & 0.2164 & 0.07878 \\
Optimal for Q\^{}(3) & 0.1220 & 0.1650 & 0.4355 & 0.2774 & 0.09035 \\
Robust minimax & \textbf{0.0953} & \textbf{0.4053} & \textbf{0.2714} &
\textbf{0.2279} & \textbf{0.07341} \\
\end{longtable}

The least-favourable target is approximately
\(0.0001Q^{(1)} + 0.7201Q^{(2)} + 0.2798Q^{(3)}\); numerically,
\(Q^{(1)}\) is essentially inactive at the minimax optimum. The robust
design equalizes the active worst-case widths and improves the maximum
width over all three single-target designs. This example also shows why
target uncertainty is not equivalent to simply using uniform weights.

\subsection{Five-system simultaneous
identification}\label{five-system-simultaneous-identification}

To evaluate the headline multi-system rule, we simulate \(K = 5\)
systems with stratum-specific Bernoulli success probabilities. The
target means are

\[(0.803,0.723,0.683,0.643,0.603),\]

so system 1 is uniquely best and its nearest gap is 0.08. On each
selected stratum visit, all five systems are evaluated, producing paired
differences for all \(H = 10\) pairs. We use the range-only construction
with local error \(\alpha/(HJ)\) and 2,000 Monte Carlo runs. System
outcomes are generated independently conditional on stratum; the purpose
of this experiment is to exercise simultaneous inference and
multiplicity rather than reproduce realistic benchmark covariance. The
pairwise statistics are nevertheless correlated because they share
system-level counts, and the familywise guarantee does not require
pairwise independence. Because all five systems are evaluated per
selected condition, one evaluation unit corresponds to five model calls
in this experiment.

\textbf{Table 7. Five-system simultaneous anytime-valid best-system
identification (evaluation units)}

\begin{longtable}[]{@{}
  >{\raggedright\arraybackslash}p{(\columnwidth - 10\tabcolsep) * \real{0.1667}}
  >{\raggedright\arraybackslash}p{(\columnwidth - 10\tabcolsep) * \real{0.1667}}
  >{\raggedright\arraybackslash}p{(\columnwidth - 10\tabcolsep) * \real{0.1667}}
  >{\raggedright\arraybackslash}p{(\columnwidth - 10\tabcolsep) * \real{0.1667}}
  >{\raggedright\arraybackslash}p{(\columnwidth - 10\tabcolsep) * \real{0.1667}}
  >{\raggedright\arraybackslash}p{(\columnwidth - 10\tabcolsep) * \real{0.1667}}@{}}
\toprule\noalign{}
\begin{minipage}[b]{\linewidth}\raggedright
\textbf{Design}
\end{minipage} & \begin{minipage}[b]{\linewidth}\raggedright
\textbf{Median}
\end{minipage} & \begin{minipage}[b]{\linewidth}\raggedright
\textbf{25th pct.}
\end{minipage} & \begin{minipage}[b]{\linewidth}\raggedright
\textbf{75th pct.}
\end{minipage} & \begin{minipage}[b]{\linewidth}\raggedright
\textbf{Mean}
\end{minipage} & \begin{minipage}[b]{\linewidth}\raggedright
\textbf{Errors}
\end{minipage} \\
\midrule\noalign{}
\endhead
\bottomrule\noalign{}
\endlastfoot
Uniform & 10,680 & 9,480 & 11,980 & 10,764.3 & 0/2000 \\
Q-proportional & 9,780 & 8,880 & 10,780 & 9,830.95 & 0/2000 \\
Range-width identification & \textbf{9,380} & 8,380 & 10,380 &
\textbf{9,454.75} & 0/2000 \\
\end{longtable}

Zero observed errors do not establish a zero error probability; the
theoretical statement is the familywise bound in \cref{cor:best-system}. The
simulation instead demonstrates that the multiple-system rule is
computationally operational and illustrates the price of the \(HJ\)
Bonferroni split. Stepwise e-value procedures or best-arm methods may
reduce this cost. All 2,000 runs identified a winner before the
monitoring horizon under each allocation; unresolved frequency was
therefore zero in this experiment. For the range-width design, the
median 9,380 evaluation units correspond to 46,900 model calls under the
five-calls-per-unit accounting.

\subsection{Why ordinary repeated Wald intervals are
unsafe}\label{why-ordinary-repeated-wald-intervals-are-unsafe}

Finally, we set all pairwise gaps to zero and repeatedly inspect an
ordinary two-sided 95\% Wald interval after each batch, stopping
whenever its lower endpoint exceeds zero. Across 10,000 null
simulations, this repeated-look rule had a 16.38\% false-positive rate,
compared with the 2.5\% one-sided tail that would apply to a single
preplanned inspection of the same two-sided interval. Under the same
monitoring horizon, the conservative range-only anytime-valid
construction produced 0 false positive identifications in 10,000
simulations. The latter count is only a Monte Carlo observation; the
inferential guarantee comes from time-uniform coverage rather than from
the observed zero.

For scale, a variance-known fixed-time one-sided test using the Neyman
variance formula would require approximately 176 paired evaluation units
for 50\% power, 403 for 80\% power, and 558 for 90\% power at the
simulated \(\Delta_{Q} = 0.073\). These numbers are not directly
comparable to the sequential identification times because the guarantees
differ: fixed-time power, anytime validity, stratum-wise multiplicity,
and multi-system familywise identification solve different inferential
problems. Reporting this decomposition is preferable to treating every
additional evaluation unit as an undifferentiated ``cost of anytime
validity.'' This distinction parallels the classical group-sequential
literature, which controls repeated analyses by calibrating the
monitoring rule rather than repeatedly applying fixed-sample tests
(Pocock 1977; O\textquotesingle Brien and Fleming 1979).

\section{Relation to sequential inference, adaptive allocation, and
selection}\label{relation-to-sequential-inference-adaptive-allocation-and-selection}

\subsection{Stratified anytime-valid
inference}\label{stratified-anytime-valid-inference}

Turner and Grünwald (2023) develop safe sequential testing and
anytime-valid confidence sequences for stratified count data, including
compound effects across subpopulations. Their work establishes that
stratified anytime-valid inference can be built directly from
e-variables and can exploit cross-stratum structure. Wang and Ramdas
(2025) provide complementary modern confidence-sequence methodology for
Gaussian means with unknown variance. The present contribution is
therefore not the first stratified confidence sequence.

Spertus, Sridhar, and Stark (2024) are particularly close to the global
stratified-mean problem considered here. They explicitly describe the
sum of multiplicity-adjusted stratumwise confidence bounds as an
unnecessarily conservative baseline, then sharpen global inference by
representing the weighted-mean null as a union of intersection
hypotheses and multiplying stratumwise test supermartingales. Their
method retains stratified sampling while avoiding the full cost of the
local-Bonferroni/$L_1$ aggregation used as the local-guarantee comparator
in this paper. Thus, when only a single global weighted mean is required,
their construction is an important intermediate and potentially much
more efficient architecture. The present paper studies a different
design question: how the target distribution, the guarantees that must
remain available after sampling, and the resulting allocation objective
interact. Its boundary-rate allocation result pertains specifically to
the additive local-width architecture, while the direct pooled
construction and the Spertus et al. global stratified test represent
alternative architectures with different guarantees and efficiency
criteria.

The e-value perspective provides a general calculus for combining
nonnegative evidence while preserving validity (Vovk and Wang 2021).

\subsection{Adaptive stratified sampling and
allocation}\label{adaptive-stratified-sampling-and-allocation}

Classical Neyman allocation minimizes the variance of a terminal
stratified estimator under a fixed budget. When stratum variances are
not known in advance, adaptive stratified designs use early observations
to redirect later sampling; Salehi et al. (2010) study such sequential
allocation from a survey-sampling perspective. Dynamic sampling
allocation also appears in simulation design and selection problems
(Peng et al. 2016). Our boundary-rate result differs in objective:
within the stratum-resolved architecture, the criterion is the
target-weighted sum of local confidence-sequence half-widths.
Consequently, the first-order allocation depends on the rate and leading
constant of the sequential boundary, not only on terminal estimator
variance. Étoré and Jourdain (2010) provide a closely related
adaptive-stratification result in which estimated stratum variability
drives convergence toward variance-optimal allocation.

\subsection{Ranking, selection, and best-system
identification}\label{ranking-selection-and-best-system-identification}

Ranking and selection has a long history of procedures for identifying
the best among several populations, and recent work continues to develop
nonparametric formulations (Alshihry, Coolen-Maturi, and Coolen 2026).
Fixed-confidence best-arm identification provides information-theoretic
lower bounds and asymptotically efficient adaptive rules (Kaufmann,
Cappé, and Garivier 2016; Garivier and Kaufmann 2016). Section 7 uses a
simpler familywise anytime-valid construction because the primary design
variable in this paper is the allocation across strata composing a
target estimand. The resulting rule is deliberately conservative but
auditable; sharper structured selection procedures are an important
extension. Classical sequential ranking-and-selection work includes
Paulson (1964) and Kim and Nelson (2001), which explicitly couple
repeated sampling with elimination or selection decisions. Imbens et al.
(2026) formulate a related threshold-bandit problem in which the global
null states that no arm has a positive effect; they compare pooled
evidence across promising arms with time-uniform multiple testing of
individual arm means and use signal-to-noise-driven adaptive allocation,
with inference based on asymptotic central-limit approximations rather than
the nonasymptotic bounded-data sequences used here.
Sandoval, Waudby-Smith, and Jordan (2026) consider a nonasymptotic
multi-armed betting problem for rejecting a global null when at least one
arm is non-null, developing log-optimal and expected-rejection-time
optimal e-processes. In both cases the inferential target is an
existential arm-level alternative. Here the estimand is instead the
prespecified weighted average \(\Delta_Q\) across strata, and the
boundary-rate allocation criterion minimizes aggregate target width
within a specified architecture rather than maximizing arm-level evidence
or minimizing rejection time for an any-arm alternative. In the
target-set architecture the contrast is exact: a global any-arm null is an
intersection of arm nulls with a union alternative, while the robust
target-set null of \cref{rem:iut} is a union of vertex nulls with an
intersection alternative. Their expected-rejection-time optimality and our
characteristic-time results (\cref{thm:lower-bound,prop:ternary}) therefore
address dual geometries.

\subsection{Target-specific and off-policy
inference}\label{target-specific-and-off-policy-inference}

Direct target inference is closely related to off-policy confidence
sequences, where predictable logging propensities are used to estimate a
target-policy value under adaptive data collection (Karampatziakis,
Mineiro, and Ramdas 2021). Waudby-Smith et al. (2024a) give a substantially
more general anytime-valid off-policy framework for contextual bandits,
including doubly robust confidence sequences for time-varying target-policy
values, empirical variance adaptation, and procedures that do not require a
known maximal importance-weight bound. The pooled scores in Sections~6.6 and
6.12 are special cases obtained by taking the action to be the stratum and the
target policy to be the externally specified target mixture; the new question
here is how one logging proposal should be designed when several scientific
target mixtures must share the same sequential data stream. Prediction-powered inference offers another
example in which auxiliary predictions improve efficiency while a
statistical correction preserves validity (Angelopoulos et al. 2023). In
the present setting, the target distribution defines the estimand and
the proposal distribution defines where information is collected. When
direct target sampling is feasible, the pooled process avoids a
stratum-wise error split; when it is not, importance weighting provides
a direct alternative subject to positivity. Some bounded-CS implementations
also pay a finite-time price for large weights, whereas the bound-free betting
methods of Waudby-Smith et al. (2024a) show that a known finite maximal weight
is not a validity requirement.
Adaptive policy-evaluation methods likewise show how reweighting can
recover valid uncertainty statements under data-dependent assignment
(Hadad et al. 2021).

\subsection{AI model evaluation as an
application}\label{ai-model-evaluation-as-an-application}

Modern benchmark evaluation makes the target-versus-sampling distinction
especially visible. Broad evaluation frameworks emphasize heterogeneous
tasks and scenarios rather than a single undifferentiated score
(Hendrycks et al. 2021; Bommasani, Liang, and Lee 2023), while recent
work argues for explicit uncertainty quantification in model evaluation
(Miller 2024). PromptEval studies efficient evaluation over multiple
prompt templates (Polo et al. 2024), and ReliableEval treats prompt
perturbation as stochastic (Lior et al. 2025). Hsu and Shekhar (2026)
are particularly close to the pooled architecture studied here: they
build a single confidence sequence for the capability of a new LLM on a
fixed question set and design active querying rules using historical
model information. Their analysis identifies prediction mismatch and a
spiky querying distribution as mechanisms that slow confidence-sequence
shrinkage. Our target-sampling and importance-weighted constructions ask
a complementary question: when evaluation conditions form prespecified
strata with an external target mixture, should inference be pooled
directly for that mixture or retain simultaneous local guarantees for
reweighting and auditability? Work on benchmark weighting, rank
uncertainty, and adaptive benchmark selection further shows that
heterogeneity and selection can matter substantively (Siska et al. 2024;
Neuhof and Benjamini 2026a, 2026b; Xu et al. 2026). The PromptEval
analysis in Section 10 is used to illustrate these statistical design
consequences; the methodology itself applies to any sequential study
with a target-weighted estimand over heterogeneous strata.

\section{Empirical illustration: heterogeneous model evaluation with
PromptEval}\label{empirical-illustration-heterogeneous-model-evaluation-with-prompteval}

We use a public multi-prompt language-model benchmark as an empirical
illustration of the general design problem. MMLU is the multitask
benchmark introduced by Hendrycks et al. (2021). The PromptEval MMLU
correctness release contains evaluations of 15 language models under 100
prompt templates across all 57 MMLU subjects (Polo et al. 2024). The
accompanying R workflow retrieves every subject-model split through the
Hugging Face Dataset Viewer API using cached files, rate-limit-aware
retries, and public data only. Across subjects there are 14,042 MMLU
items; each subject-model matrix contains 100 prompt rows and one
correctness column per item. The analysis treats subject as the stratum
and uses the released correctness matrix as an empirical population for
sequential replay.

\subsection{Targets and ranking
stability}\label{targets-and-ranking-stability}

MMLU subject is the evaluation stratum. We prespecified three targets:
(i) an equal-subject \emph{macro} target; (ii) an \emph{item-weighted}
target proportional to the number of MMLU items per subject; and (iii) a
\emph{STEM-emphasis} target that doubles the weight of 23 prespecified
quantitative and scientific subjects before renormalization. Their
convex hull defines \(\mathcal{Q}\) for the robust analysis.

Table~8 reports the five highest-scoring systems. The complete ordering
of all 15 systems was identical under all three targets (Spearman and
Kendall rank correlations equal one). Thus target reweighting changes
absolute benchmark scores but does not change the ranking in this data
set. The leading system, Llama-3-70B-Instruct, scores 0.7969, 0.7955,
and 0.7762 under the macro, item-weighted, and STEM-emphasis targets,
respectively. Mixtral-8x7B-Instruct-v0.1 is second under all three,
giving target gaps of 0.0982, 0.1065, and 0.1051.

\textbf{Table 8. Top five PromptEval systems under three prespecified
target distributions. Ranks are identical across targets}

\begin{longtable}[]{@{}
  >{\raggedright\arraybackslash}p{(\columnwidth - 6\tabcolsep) * \real{0.2500}}
  >{\raggedright\arraybackslash}p{(\columnwidth - 6\tabcolsep) * \real{0.2500}}
  >{\raggedright\arraybackslash}p{(\columnwidth - 6\tabcolsep) * \real{0.2500}}
  >{\raggedright\arraybackslash}p{(\columnwidth - 6\tabcolsep) * \real{0.2500}}@{}}
\toprule\noalign{}
\begin{minipage}[b]{\linewidth}\raggedright
\textbf{System}
\end{minipage} & \begin{minipage}[b]{\linewidth}\raggedright
\textbf{Macro}
\end{minipage} & \begin{minipage}[b]{\linewidth}\raggedright
\textbf{Item-weighted}
\end{minipage} & \begin{minipage}[b]{\linewidth}\raggedright
\textbf{STEM-emphasis}
\end{minipage} \\
\midrule\noalign{}
\endhead
\bottomrule\noalign{}
\endlastfoot
Llama-3-70B-Instruct & 0.7969 & 0.7955 & 0.7762 \\
Mixtral-8x7B-Instruct-v0.1 & 0.6987 & 0.6890 & 0.6711 \\
Falcon-180B & 0.6807 & 0.6741 & 0.6474 \\
Llama-3-8B-Instruct & 0.6528 & 0.6416 & 0.6245 \\
Llama-3-8B & 0.6336 & 0.6230 & 0.6055 \\
\end{longtable}

For the sequential design replay we fix the two highest-ranked systems
under the prespecified macro target: Llama-3-70B-Instruct and
Mixtral-8x7B-Instruct-v0.1. This is not a claim of post-selection
inference for the original benchmark population. The public matrix is
treated as a fixed empirical population, the pair is then held fixed,
and repeated replay samples are drawn from that population to compare
prospective evaluation policies.

\subsection{Subject heterogeneity and robust target
design}\label{subject-heterogeneity-and-robust-target-design}

The pairwise advantage is positive in all 57 subjects, but its magnitude
is heterogeneous. The subject-specific mean difference ranges from
0.0029 to 0.3161, with median 0.0944; the corresponding paired standard
deviation ranges from 0.2385 to 0.6365, with median 0.4048. This
combination---stable sign but heterogeneous variance and effect
magnitude---is useful for evaluating design efficiency without relying
on a target-induced rank reversal.

The robust analysis takes
\(\mathcal{Q} = conv\{ Q_{macro},Q_{item},Q_{STEM}\}\). The
least-favourable target mixture is approximately

\[0.1720Q_{macro} + 0.1231Q_{item} + 0.7050Q_{STEM}.\]

For the first-order variance-width objective, the associated worst-case
coefficient is 3.251, so the leading width is approximately
\(3.251/\sqrt{B}\) at total budget \(B\). The resulting robust
allocation is diffuse rather than concentrated on a few subjects: its
effective number of strata, \(1/\sum_{j}^{}w_{j}^{2}\), is about 53.0 of
57, and its largest single subject share is 0.0270 (high-school
mathematics). Because the three PromptEval targets are deliberately
close perturbations of one benchmark composition, this calculation is
best viewed as an illustration of the robust-design mechanics rather
than evidence about severe deployment shift. Wider target sets would be
needed to study genuinely adversarial composition uncertainty.

\subsection{Sequential replay and evaluation
cost}\label{sequential-replay-and-evaluation-cost}

A replay evaluation unit is one prompt--item cell sampled with
replacement from a selected MMLU subject, with the same cell used for
both systems. This treats the released matrix as an empirical population
and matches the within-stratum i.i.d. local-time assumptions used by the
current theory. We use 500 Monte Carlo replay trajectories, five pilot
observations per subject, batches of 100, \(\alpha = .05\), and a
maximum of 100,000 paired evaluation units. The adaptive stratified
policy uses \(\varepsilon_{t} = \min\{ 0.25,t^{- 0.25}\}\) with
exploration distribution \(\nu = Q\). Every policy identified the
positive direction in every trajectory; there were no unresolved or
negative identifications. Table~9 therefore resamples observed
prompt--item cells. For Table~10, the pooled sampler instead
reconstructs each subject's exact ternary distribution on
\(\{ - 1,0,1\}\) from its empirical mean \(\delta_{j}\) and standard
deviation \(\sigma_{j}\): because \(D^{2}\) is binary on this support,
\(E\left( D^{2} \right) = \sigma_{j}^{2} + \delta_{j}^{2}\) together
with \(E(D) = \delta_{j}\) uniquely determines the probabilities of
\(- 1\), \(0\), and \(1\). Thus the two tables use different sampling
implementations but the pooled reconstruction preserves exactly the
subject-level first two moments and, for a ternary paired difference,
the full empirical distribution.

Table~9 gives the central empirical comparison. Under the macro target,
most fixed policies require roughly 32,000--33,000 paired evaluation
units, while adaptive PrPl-EB reduces the median from 32,485 under
uniform sampling to 30,385, a 6.5\% saving. Under item weighting the
design choice matters much more: uniform sampling requires a median
33,085 units and \(Q\)-proportional sampling 25,485, so merely aligning
the proposal with the target accounts for a 23.0\% reduction. Moving
from \(Q\)-proportional to the range-width design lowers the median
further to 23,985, an additional 5.9\% reduction and 27.5\% relative to
uniform. The adaptive design has the same median 23,985. Neyman
allocation, despite being estimation-optimal for its own objective,
requires 32,585 units and therefore provides almost no
sequential-identification saving in this target. Under STEM emphasis,
adaptive PrPl-EB gives the lowest median, 27,585 versus 29,285 under
uniform, a 5.8\% saving.

\textbf{Table 9. PromptEval paired evaluation units to PrPl-EB
identification of the leading macro pair (500 replay trajectories)}

\begin{longtable}[]{@{}
  >{\raggedright\arraybackslash}p{(\columnwidth - 10\tabcolsep) * \real{0.1667}}
  >{\raggedright\arraybackslash}p{(\columnwidth - 10\tabcolsep) * \real{0.1667}}
  >{\raggedright\arraybackslash}p{(\columnwidth - 10\tabcolsep) * \real{0.1667}}
  >{\raggedright\arraybackslash}p{(\columnwidth - 10\tabcolsep) * \real{0.1667}}
  >{\raggedright\arraybackslash}p{(\columnwidth - 10\tabcolsep) * \real{0.1667}}
  >{\raggedright\arraybackslash}p{(\columnwidth - 10\tabcolsep) * \real{0.1667}}@{}}
\toprule\noalign{}
\begin{minipage}[b]{\linewidth}\raggedright
\textbf{Target}
\end{minipage} & \begin{minipage}[b]{\linewidth}\raggedright
\textbf{Policy}
\end{minipage} & \begin{minipage}[b]{\linewidth}\raggedright
\textbf{Median}
\end{minipage} & \begin{minipage}[b]{\linewidth}\raggedright
\textbf{25th pct.}
\end{minipage} & \begin{minipage}[b]{\linewidth}\raggedright
\textbf{75th pct.}
\end{minipage} & \begin{minipage}[b]{\linewidth}\raggedright
\textbf{Mean}
\end{minipage} \\
\midrule\noalign{}
\endhead
\bottomrule\noalign{}
\endlastfoot
Macro & Uniform & 32,485 & 31,385 & 33,385 & 32,456.4 \\
& Q-proportional & 32,335 & 31,485 & 33,285 & 32,321.2 \\
& Neyman estimation & 32,785 & 31,985 & 33,685 & 32,852.6 \\
& Range-width & 32,485 & 31,485 & 33,385 & 32,445.2 \\
& First-order variance-optimal & 32,385 & 31,585 & 33,285 & 32,410.4 \\
& Adaptive PrPl-EB & \textbf{30,385} & \textbf{29,385} & \textbf{31,185}
& \textbf{30,363.0} \\
Item & Uniform & 33,085 & 31,385 & 34,585 & 33,077.0 \\
& Q-proportional & 25,485 & 24,785 & 26,185 & 25,512.6 \\
& Neyman estimation & 32,585 & 31,585 & 33,685 & 32,635.4 \\
& Range-width & \textbf{23,985} & \textbf{23,185} & 24,885 &
\textbf{24,010.6} \\
& First-order variance-optimal & 24,585 & 23,885 & 25,485 & 24,626.4 \\
& Adaptive PrPl-EB & \textbf{23,985} & 23,285 & \textbf{24,710} &
24,025.2 \\
STEM & Uniform & 29,285 & 28,185 & 30,185 & 29,163.2 \\
& Q-proportional & 29,885 & 29,085 & 30,785 & 29,963.8 \\
& Neyman estimation & 29,685 & 29,085 & 30,485 & 29,745.2 \\
& Range-width & 28,685 & 27,585 & 29,610 & 28,666.6 \\
& First-order variance-optimal & 29,285 & 28,485 & 30,085 & 29,290.4 \\
& Adaptive PrPl-EB & \textbf{27,585} & \textbf{26,785} & \textbf{28,485}
& \textbf{27,630.8} \\
\end{longtable}

\textbf{Table 10. PromptEval comparison of direct pooled inference with the
best local-CS stratified comparator for the same leading pair. Direct pooled rows use one PrPl-EB confidence
sequence at level $\alpha=.05$, five pooled pilot observations, batch-100
monitoring, and 5,000 Monte Carlo trajectories. Pooled entries are
medians {[}25th, 75th percentiles{]}}

\begin{longtable}[]{@{}
  >{\raggedright\arraybackslash}p{(\columnwidth - 8\tabcolsep) * \real{0.2000}}
  >{\raggedright\arraybackslash}p{(\columnwidth - 8\tabcolsep) * \real{0.2000}}
  >{\raggedright\arraybackslash}p{(\columnwidth - 8\tabcolsep) * \real{0.2000}}
  >{\raggedright\arraybackslash}p{(\columnwidth - 8\tabcolsep) * \real{0.2000}}
  >{\raggedright\arraybackslash}p{(\columnwidth - 8\tabcolsep) * \real{0.2000}}@{}}
\toprule\noalign{}
\begin{minipage}[b]{\linewidth}\raggedright
\textbf{Target}
\end{minipage} & \begin{minipage}[b]{\linewidth}\raggedright
\textbf{Best local-CS strat.}
\end{minipage} & \begin{minipage}[b]{\linewidth}\raggedright
\textbf{Pooled $\pi = Q$}
\end{minipage} & \begin{minipage}[b]{\linewidth}\raggedright
\textbf{Pooled 2nd-moment}
\end{minipage} & \begin{minipage}[b]{\linewidth}\raggedright
\textbf{Saving vs. local-CS}
\end{minipage} \\
\midrule\noalign{}
\endhead
\bottomrule\noalign{}
\endlastfoot
Macro & 30,385 & \textbf{305 {[}205, 405{]}} & 405 {[}305, 505{]} &
99.0\% \\
Item & 23,985 & \textbf{305 {[}205, 405{]}} & 405 {[}305, 405{]} &
98.7\% \\
STEM & 27,585 & \textbf{305 {[}205, 405{]}} & 405 {[}305, 405{]} &
98.9\% \\
\end{longtable}

Table~10 changes the hierarchy of conclusions. The 6--27\% differences
among stratified allocation rules are real, but they are second-order
relative to the cost of requiring 57 simultaneous local confidence
sequences in this construction. Sampling directly from the prespecified
target \(Q\) makes each observed paired difference itself a bounded
unbiased observation of \(\Delta_{Q}\), so the pooled procedure needs
only one confidence sequence and, under the same batch-100 monitoring
granularity used in Table~9, stops at a median of 305 evaluation units
for each of the macro, item, and STEM targets. The direct second-moment
proposal in Equation~\eqref{eq:3.5} reduces asymptotic variance but has
importance-weight bound \(B(\pi) \approx 1.83\)--\(1.90\) in PromptEval
and is slower at finite sample here, with a median of 405 units for each
target. Batch-10 monitoring gives the same qualitative conclusion with
target-sampling medians of 255, 235, and 235 and second-moment medians
of 345, 315, and 325, respectively. Together with the synthetic result
in Section~8.5, where the second-moment proposal is faster, this shows
that proposal spikiness is a context-dependent finite-sample cost rather
than a general dominance result.

Figure 2 visualizes the architecture comparison from Table 10; the stratified bar refers specifically to the simultaneous local-CS comparator.

\includegraphics[width=6in,height=3.29448in]{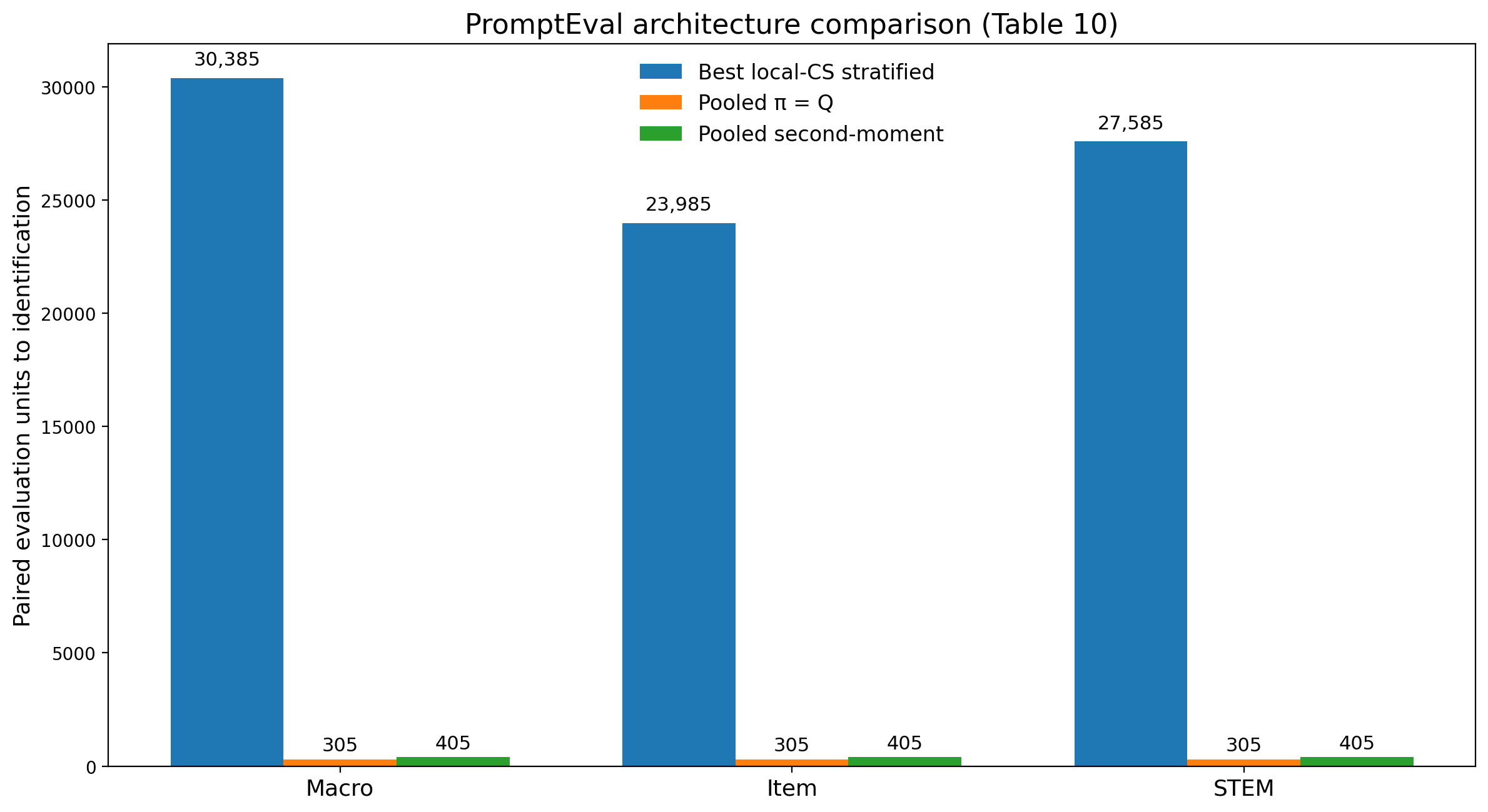}

\emph{Figure 2. PromptEval replay under the bounded PrPl-EB implementation. Direct pooled confidence sequences require fewer paired evaluation units than the simultaneous local-CS aggregation baseline, which maintains 57 local guarantees and is intentionally conservative when only a global target is required. The figure is not a comparison against sharper global stratified procedures such as the combined-supermartingale construction of Spertus, Sridhar, and Stark (2024).}

Two aspects of Table~9 matter \emph{within the stratum-resolved
architecture}. First, the empirically best allocation depends on the
target and finite-sample boundary; the first-order
\(\left( q_{j}\sigma_{j} \right)^{2/3}\) allocation is an asymptotic
width optimum, not a claim of exact finite-time optimality for PrPl-EB.
The adaptive policy can therefore stop earlier than this static
first-order benchmark on some finite replay paths because it responds to
realized variance estimates and boundary geometry; that observation is
not evidence of beating a finite-time optimum. Second, the item-weighted
target gives a real-data analogue of the paper's central objective
distinction: a design tailored to fixed-budget estimation (Neyman) is
much less effective for sequential identification than
identification-oriented sampling. The stratified empirical finding is
therefore one of \emph{cost sensitivity under rank stability}: the
target does not reverse the winner here, but it changes where evaluation
effort should be spent. The pooled comparison adds a more consequential
lesson: if only the target-level sign is required, the choice between
pooled and stratum-resolved inference dominates the smaller differences
among stratified allocations.

\subsection{One shared stream for a prespecified target set}\label{target-set-replay}

The three PromptEval targets also provide a direct test of the
intermediate architecture in Section~6.7. Rather than running a separate
pooled experiment for each target, we draw one sequence of subjects from
a common proposal and update three importance-weighted pooled confidence
sequences simultaneously, each at level \(1-\alpha/3\). The procedure
stops only when all three lower bounds are positive (or all three upper
bounds are negative), which certifies the sign uniformly over
\(\operatorname{conv}\{Q_{macro},Q_{item},Q_{STEM}\}\).

For these targets, the intrinsic target-set range factor is
\(C_{\mathcal Q}=1.4324\), the smallest achievable worst importance-weight bound for any shared proposal. The unconstrained minimax second-moment proposal
places essentially all least-favourable dual weight on the item and STEM
targets, with \(\lambda^*\approx(0.0000,0.4420,0.5580)\). Its worst
second-moment proxy is 0.24814, but its largest target-specific
importance-weight bound is 2.835. Range capping sacrifices a small
amount of second-moment efficiency to reduce this bound.

\textbf{Table 11. PromptEval simultaneous target-set pooled inference under the bounded rescaled PrPl-EB implementation. One shared sampling stream updates three two-sided target-level confidence sequences at error level \(\alpha/3\). The static proposals use population moments from the full empirical replay population and are therefore population-informed (oracle) benchmarks. 5,000 trajectories, five pilot observations, and batch-10 monitoring. Mean stopping times are followed by Monte Carlo standard errors in parentheses.}

{\small
\begin{longtable}[]{@{}
  >{\raggedright\arraybackslash}p{0.29\columnwidth}
  >{\raggedright\arraybackslash}p{0.11\columnwidth}
  >{\raggedright\arraybackslash}p{0.10\columnwidth}
  >{\raggedright\arraybackslash}p{0.19\columnwidth}
  >{\raggedright\arraybackslash}p{0.18\columnwidth}@{}}
\toprule
\textbf{Shared proposal} & \textbf{Worst }\(S_m\) & \textbf{Max }\(B_m\) & \textbf{Median [IQR]} & \textbf{Mean (MCSE)} \\
\midrule
\endhead
\bottomrule
\endlastfoot
Average target & 0.29871 & 2.353 & 545 [455, 675] & 576.49 (2.36) \\
Population-informed minimax second-moment & \textbf{0.24814} & 2.835 & 645 [555, 755] & 663.22 (2.05) \\
Population-informed range-capped, \(C=C_{\mathcal Q}=1.432\) & 0.27344 & \textbf{1.432} & \textbf{445 [355, 575]} & \textbf{475.51 (2.39)} \\
Population-informed range-capped, \(C=1.50\) & 0.25975 & 1.500 & 455 [365, 565] & 478.75 (2.28) \\
\end{longtable}
}

Table~11 is a comparison within one specific finite-sample confidence-sequence implementation. The unconstrained minimax proposal is optimal for the worst second-moment proxy but is slower under the bounded rescaled PrPl-EB sequence because its rescaling constants are larger. The range-capped proposals are faster in this implementation, but this ordering should not be interpreted as a universal penalty for uncapped off-policy inference: bound-free betting confidence sequences do not require a known maximal importance weight (Waudby-Smith et al. 2024a). All 5,000 trajectories identified a positive effect uniformly across the three target vertices. The replay draws from exact ternary distributions reconstructed from each subject's empirical \((\delta_j,\sigma_j)\), as in Table~10, rather than directly resampling prompt--item cells.

For robust directional certification we use the classical intersection--union logic of \cref{rem:iut}, with a one-sided lower PrPl-EB sequence for each target vertex (the \(\log(1/\alpha)\) rather than \(\log(2/\alpha)\) boundary term). Each vertex process is run at level \(\alpha=.05\), with no \(\alpha/M\) split, and rejection occurs only when every lower bound is positive. The same reconstructed ternary replay population and batch-10 monitoring are used. We compare the population-informed second-moment and gap-aware proposals of Sections~6.8--6.10.

\textbf{Table 12. PromptEval robust-positive target-set certification using one-sided lower PrPl-EB sequences at nominal level \(\alpha=.05\). One shared sampling stream; rejection requires all three lower bounds to exceed zero. Static proposals use full-population gaps and moments and are therefore population-informed (oracle) benchmarks. 5,000 trajectories, five pilot observations, batch-10 monitoring. Mean stopping times are followed by Monte Carlo standard errors in parentheses.}

{\small
\begin{longtable}[]{@{}
  >{\raggedright\arraybackslash}p{0.29\columnwidth}
  >{\raggedright\arraybackslash}p{0.12\columnwidth}
  >{\raggedright\arraybackslash}p{0.09\columnwidth}
  >{\raggedright\arraybackslash}p{0.18\columnwidth}
  >{\raggedright\arraybackslash}p{0.15\columnwidth}@{}}
\toprule
\textbf{Shared proposal} & \textbf{Worst gap-scaled proxy} & \textbf{Max }\(B_m\) & \textbf{Median [IQR]} & \textbf{Mean (MCSE)} \\
\midrule
\endhead
\bottomrule
\endlastfoot
Population-informed minimax second-moment & 23.660 & 2.835 & 405 [345, 485] & 422.77 (1.62) \\
Population-informed gap-aware minimax & \textbf{22.373} & 2.444 & 365 [305, 445] & 384.38 (1.59) \\
Population-informed gap-aware, range-capped \(C=C_{\mathcal Q}=1.432\) & 24.751 & \textbf{1.432} & \textbf{285 [205, 385]} & \textbf{312.19 (2.00)} \\
Population-informed gap-aware, range-capped \(C=1.50\) & 23.436 & 1.500 & \textbf{285 [215, 375]} & \textbf{310.62 (1.83)} \\
\end{longtable}
}

Table~12 isolates the directional guarantee: there is no \(M\)-fold split and no two-sided factor. Incorporating target gaps improves the uncapped proposal, and range capping gives the smallest finite-sample stopping times under the bounded rescaled PrPl-EB implementation; the two capped designs are indistinguishable at this Monte Carlo resolution (mean difference 1.6 with standard errors near 2). The latter is a property of this operational confidence sequence, not a universal theorem about off-policy inference. The static proposals in Tables~11--12 use population quantities and therefore benchmark the design objectives rather than implementing the adaptive plug-in policy of \cref{thm:gap-tracking}; a fully adaptive replay is left for subsequent work.

The batch-10 replays also give a matched guarantee--cost ladder for the same leading pair. A single prespecified target, monitored by one one-sided lower PrPl-EB sequence at level \(\alpha\) under target sampling, stops at medians of 205, 185, and 185 evaluation units for the macro, item, and STEM targets (means 227.3, 209.0, and 214.2; standard errors 1.5--1.7). Certifying the sign uniformly over their convex hull with one shared stream (Table~12, capped designs) costs a median of 285, about 80--100 additional units. Simultaneous two-sided reporting of all three target gaps at level \(\alpha/3\) (Table~11) costs a median of 445, and the local-CS comparator with 57 stratum sequences costs more than 23,000 (Table~9). Table~10 uses batch-100 monitoring and two-sided sequences, so its medians should not be compared directly with Tables~11--12.

\subsection{Information-limit benchmark for the PromptEval target set}\label{information-benchmark}

To connect the replay to \cref{thm:lower-bound}, we form a Gaussian benchmark with the same
three target gaps and the same 57 within-subject standard deviations. This is
not a finite-sample approximation to the bounded replay; it is an
information-theoretic calibration of the target-set geometry. Table~13 reports
the single-target characteristic times and the robust three-target optimum. For
one target, Equation~\eqref{eq:6.30} gives the Neyman-type centered allocation. For all
three targets jointly, Equation~\eqref{eq:6.28} equalizes the active target-specific
variance-to-gap criteria.

\textbf{Table 13. Information-limit benchmarks using the PromptEval target gaps and reconstructed subject-level distributions. Gaussian columns use the known-variance model in \cref{thm:lower-bound}; ternary columns use the exact characteristic information in \cref{prop:ternary}, which by \cref{lem:klinf} equals the bounded nonparametric (\(\mathrm{KL}_{\inf}\)) information for this population. The lower-bound columns report \(\log(20)/\Gamma\) for an almost-surely successful level-.05 robust-positive procedure. These are information benchmarks, not predictions of the finite-sample PrPl-EB replay.}

{\small
\begin{longtable}[]{@{}
  >{\raggedright\arraybackslash}p{0.17\columnwidth}
  >{\raggedright\arraybackslash}p{0.08\columnwidth}
  >{\raggedright\arraybackslash}p{0.13\columnwidth}
  >{\raggedright\arraybackslash}p{0.14\columnwidth}
  >{\raggedright\arraybackslash}p{0.13\columnwidth}
  >{\raggedright\arraybackslash}p{0.14\columnwidth}@{}}
\toprule
\textbf{Problem} & \textbf{Gap} & \(\Gamma_G^*\) & \(\log(20)/\Gamma_G^*\) & \(\Gamma_{ter}^*\) & \(\log(20)/\Gamma_{ter}^*\) \\
\midrule
\endhead
\bottomrule
\endlastfoot
Macro target & 0.0982 & 0.02608 & 114.9 & 0.02562 & 116.9 \\
Item target & 0.1065 & 0.02920 & 102.6 & 0.02842 & 105.4 \\
STEM target & 0.1051 & 0.02777 & 107.9 & 0.02728 & 109.8 \\
All three targets & --- & \textbf{0.02383} & \textbf{125.7} & \textbf{0.02300} & \textbf{130.2} \\
\end{longtable}
}

The ternary benchmark, which by \cref{lem:klinf} is the bounded
nonparametric limit for this population, is slightly more demanding than the
Gaussian surrogate for each target and for the joint three-target problem.  At
the ternary optimum all three target vertices are active to numerical tolerance;
the optimal shared proposal has an effective number of about 42.9 of the 57
strata.  The level-.05 robust lower bound is 130.2 observations, compared with
125.7 under the Gaussian surrogate.  Both remain below the bounded PrPl-EB
stopping times in Tables~11--12, as expected: the information bounds abstract
away from the particular operational confidence sequence, batch overshoot, and
tuning.  Their role is to identify the irreducible target-set difficulty and the
allocation geometry, not to approximate the replay stopping time directly.

The target-set experiments are reproduced by
\path{reproduce_targetset_replay.py} (Tables~11--12 and the matched
single-target baseline), \path{reproduce_information_bound_v2.py} (Gaussian
columns of Table~13), and \path{reproduce_ternary_information_bound.py}
(ternary columns of Table~13); the shipped outputs include the batch-10
simultaneous, robust-positive, and single-target replay summaries, both
information benchmarks, and the corresponding proposal files.

The retrieval, summaries, target rankings, subject-level pair
diagnostics, robust allocation, and sequential replay are generated by
\path{prompteval_target_aware_replay_rate_safe.R}. Compact
CSV/RDS outputs are included with the reproducibility materials; the
public correctness matrix remains distributed through PromptEval's
original channel.

\section{Design recommendations}\label{design-recommendations}

The framework yields several operational recommendations for sequential
studies with heterogeneous strata.

First, state the target distribution before optimizing the evaluation. A
uniformly sampled benchmark answers a uniform-target question only if
uniform weighting is scientifically intended. If deployment weights are
available, the estimand should be written explicitly as \(\Delta_{Q}\).

Second, choose the inferential guarantee before choosing the architecture
or allocation. If only one prespecified target \(Q\) is needed, one pooled
bounded-mean confidence sequence is a strong default. If a finite set of
target mixtures is prespecified but local stratum intervals are not
needed, the target-set pooled architecture of Section~6.7 provides an
intermediate option. Within that architecture, distinguish simultaneous
target-wise reporting from robust directional certification: the classical
intersection--union argument in Section~6.9 shows that the latter does not
require an M-fold error split. Full stratum-resolved inference should be
reserved for local reporting, unrestricted post-hoc reweighting, or
guarantees that cannot be represented by the prespecified target family.
Only then optimize the proposal or allocation within the selected
guarantee class.

Third, use variance-adaptive bounded-mean confidence sequences rather
than range-only sequences when evaluation metrics permit. The
simulations show that boundary efficiency can dominate the gains from
fine-tuning allocation. When pooled inference is used, predictable centering
in \cref{prop:centering} removes first-order between-stratum mean heterogeneity and
provides a principled route toward the information-optimal allocation in
\cref{cor:info-optimal}. The Gaussian characteristic time in \cref{thm:lower-bound} is useful as a
benchmark for whether a proposed design is spending observations on the right
strata even when the operational confidence sequence is non-Gaussian and
finite-sample valid.

Fourth, preserve exploration while learning the design. \cref{thm:tracking}
shows that a vanishing exploration floor can learn the local width-optimal
allocation, while \cref{thm:gap-tracking} gives the corresponding result for the
gap-aware shared proposal in target-set pooled inference. In both cases,
\(\sum_t\varepsilon_t=\infty\) guarantees continuing information and
\(\varepsilon_t\downarrow0\) removes the exploration mixture
asymptotically. If direct off-policy inference is used, overlap still
matters because very small proposal probabilities create unstable
importance weights.

Fifth, separate exploratory search from confirmatory identification when
targets, outcomes, model families, or stopping rules have been tuned
after inspecting the data. Simultaneous confidence sequences protect
against selecting a current leader within a prespecified family, but
they do not erase bias created by unrestricted upstream search. A
held-out confirmatory stage or a prespecified evaluation protocol
remains appropriate when extensive exploratory tuning precedes the
sequential analysis.

\section{Limitations and
extensions}\label{limitations-and-extensions}

Several limitations are important.

\subsection{Empirical replay rather than a prospective
study}\label{empirical-replay-rather-than-a-prospective-study}

The public-data analysis is an empirical replay, not a prospective
benchmark study. It treats the released PromptEval correctness matrix as
a finite empirical population and samples prompt-item cells with
replacement. The selected pair is fixed after inspection of the full
macro ranking, so the replay compares evaluation cost conditional on
that pair; it is not post-selection inference about an unknown
superpopulation. The 98.7--99.0\% reduction in median stopping cost
under the batch-100 comparison is therefore not a universal constant: it
reflects 57 strata, the chosen target gaps, boundary construction, error
allocation, and monitoring schedule. \cref{prop:architecture} explains the
direction of the architecture penalty in a symmetric benchmark, while
the replay quantifies it for this particular empirical population. A
prospective study should prespecify systems or use held-out or
selection-aware confirmation when the candidate set is itself chosen
adaptively.

\subsection{Dependence within
strata}\label{dependence-within-strata}

The simple allocation derivations assume independent stratum
observations. Real benchmark items can be correlated, and repeated
prompt templates may share content or examples. In the fixed-budget
setting the variance becomes a quadratic form rather than
Equation~\eqref{eq:4.1}. Sequential validity likewise requires martingale
assumptions appropriate to the dependence structure. Blocked or
clustered e-processes are possible directions, but they require separate
development.

\subsection{Variance-adaptive
tuning}\label{variance-adaptive-tuning}

The PrPl-EB construction is finite-sample valid for any predictable
betting sequence in its admissible range, but its efficiency depends on
the predictable variance estimate and tuning cap. The present
implementation uses a simple stabilized plug-in. More aggressive betting
constructions may be tighter. Moreover, the variance-sensitive \(2/3\)
law is a first-order design result; exact finite-time optimal allocation
for a specific betting boundary remains a numerical optimization
problem.

\subsection{Multiplicity with many
systems}\label{multiplicity-with-many-systems}

The \(HJ\) Bonferroni split is easy to audit but can become expensive.
Structured e-value aggregation, stepwise procedures, and BAI-inspired
stopping rules could provide tighter simultaneous identification.
Characterizing the optimal allocation when one selected evaluation
condition yields a vector of outcomes across many candidate systems is
particularly interesting.

The architecture comparison should also be interpreted according to the
guarantees being purchased. The stratum-resolved comparator used here
maintains simultaneous local confidence sequences and aggregates their
widths, which is appropriate when local reporting, post-hoc reweighting,
or target-set robustness is required. It is intentionally conservative
if the sole goal is one prespecified target. Intermediate constructions can preserve stratified sampling without
paying the full local Bonferroni and \(L_{1}\)-aggregation cost. A
concrete example is the union-of-intersections construction of Spertus,
Sridhar, and Stark (2024), which combines stratumwise test
supermartingales to test the global weighted mean directly. A single
confidence sequence for a stratified target estimator is another
possibility. The large PromptEval reductions therefore quantify the cost
of this particular simultaneous-local-guarantee architecture, not a
universal advantage of every pooled method over every stratified method.
A direct numerical comparison with sharper global stratified procedures
is an important next benchmark.

\subsection{Prespecification and multiplicity in target-set pooled inference}\label{targetset-limitations}

The target-set pooled architecture requires the target vertices to be
specified before monitoring, or otherwise protected by an appropriate
selection adjustment. Simultaneous target-wise confidence reporting pays
multiplicity across the \(M\) target sequences, although this can still
be far cheaper than maintaining \(J\) local sequences when \(M\ll J\).
For the weaker decision that every target has the same direction,
the intersection--union argument in Section~6.9 removes the M-fold split; it does not provide simultaneous nominal-level intervals for
all target means. \cref{prop:minimax-proposal} optimizes a raw second-moment criterion,
whereas \cref{prop:gap-aware} adds a gap-aware first-order stopping proxy.
Neither criterion alone is exact for a bounded finite-time confidence
sequence because importance-weight range also matters. The cap in
\cref{cor:waterfill} exposes that axis, and Table~12 shows that a capped design
can outperform the uncapped first-order optimum at finite sample.
Optimal data-adaptive tuning of the cap while preserving a predeclared
guarantee remains open. Finally, target-set pooled inference does not
recover individual stratum effects and therefore cannot substitute for
local confidence sequences when stratum-level auditability is itself a
scientific requirement.

\subsection{Scope of the information-optimality benchmark}\label{information-limitations}

\cref{thm:lower-bound} is an information-theoretic benchmark for independent Gaussian
strata with known variances. \cref{prop:ternary} and \cref{lem:klinf} complement it with the exact
bounded nonparametric (\(\mathrm{KL}_{\inf}\)) characteristic information for
the reconstructed PromptEval population, so the information calibration is not
solely Gaussian. Neither
benchmark claims that the bounded PrPl-EB implementation in Tables~10--12
attains the corresponding finite-sample constant. \cref{prop:attainment} is deliberately
conditional on an asymptotically sharp target-level boundary and allocation
tracking; proving a matching constant for a fully specified finite-sample-valid
bounded-data betting procedure, including second-order effects, remains open. Likewise, predictable centering
preserves unbiasedness exactly, but finite-time confidence-sequence efficiency
depends on how the residual process is studentized or bounded.

\subsection{Construction of the target-uncertainty
set}\label{construction-of-the-target-uncertainty-set}

The robust extension treats the set of plausible targets as
scientifically supplied. It does not solve the upstream problem of
estimating deployment weights or deciding how large the uncertainty set
should be. When \(Q\) is learned from finite deployment data, a fuller
analysis should propagate that estimation uncertainty into
\(\mathcal{Q}\) or jointly model the target and evaluation streams.

\section{Discussion}\label{discussion}

The central statistical point is that target definition, data
collection, and sequential inference are separate design layers. The
target distribution specifies the population quantity of interest; the
sampling policy determines where limited observations are collected; and
the inferential architecture determines which uncertainty statements
must remain valid under continuous monitoring. Treating these layers as
interchangeable obscures both the estimand and the relevant notion of
efficiency.

The main example is Neyman allocation. It is optimal for a well-defined
fixed-budget variance objective, but that does not make it optimal for
an anytime-valid stopping rule. A range-only confidence sequence has a
different geometry, leading to a \(q^{2/3}\) allocation at first order.
More generally, \cref{prop:boundary-rate} shows that the allocation exponent is
determined by the boundary rate. A variance-adaptive root-\(n\) boundary
changes the relevant coefficient and yields \((q\sigma)^{2/3}\) under
equal costs, but still not \(q\sigma\). \cref{thm:tracking} further shows that
this optimal allocation can be learned online with a vanishing
exploration fraction. Thus ``optimal evaluation design'' has no unique
meaning until the inferential objective is specified.

This distinction also clarifies an apparent paradox in the simulations.
\(Q\)-proportional sampling is worse than uniform sampling for
fixed-budget estimator variance in the chosen heterogeneous DGP, yet
better than uniform sampling for range-only sequential identification.
Neither result is contradictory. The procedures optimize different
functions of the allocation.

The second major lesson is stronger than boundary tuning:
confidence-sequence architecture can dominate allocation. The
stratum-resolved construction pays for simultaneous local guarantees
through additive width and a local error split. \cref{prop:architecture} makes
this cost explicit in a symmetric root-n benchmark, where the optimized
stratified half-width carries a square-root-of-J first-order penalty
before accounting for the stricter local boundary constant. If local
guarantees are scientifically required, variance-adaptive betting and
the boundary-rate law provide principled ways to allocate samples within
that architecture. If they are not required, pooled target-specific
inference can avoid much of the cost entirely. This contrast is not a
claim that pooled inference dominates all stratified procedures: Spertus,
Sridhar, and Stark (2024), for example, show how a global stratified test
can combine evidence across strata far more efficiently than separate
local confidence bounds when local intervals themselves are not required.

The target-set pooled construction adds a middle point to this architecture
spectrum. When several target mixtures are scientifically relevant but
stratum-level intervals are not, simultaneous target-level pooled
processes preserve exactly the prespecified reweighting flexibility. The
minimax shared proposal then makes the sampling distribution part of the
same robust-design problem. \cref{prop:minimax-proposal} shows that the least-favourable
target combination enters quadratically through the target weights,
whereas \cref{prop:least-favourable} for the local-CS width problem uses a linear
least-favourable target mixture. \cref{prop:min-range} identifies the exact
minimum shared importance-weight range as a geometric property of the target
set. The classical intersection--union argument in Section~6.9 adds a second
distinction inside the target-set architecture itself: simultaneous reporting
of all target means and certification of a common direction are different
multiple-inference problems, and the latter avoids an M-fold error split. \cref{prop:gap-aware} then makes the proposal decision-sensitive
by scaling target-specific second moments by inverse squared target gaps,
and \cref{thm:gap-tracking} shows that this shared proposal can be learned online
under vanishing exploration. The target-set architecture therefore has
its own validity, robust-design, stopping-scale, range-control, and
adaptive-learning theory rather than serving only as an interpolation
between the two original architectures.

The information-theoretic analysis adds a further layer. Predictable centering
shows that the raw second-moment penalty is not fundamental: once local mean
heterogeneity is removed by a past-measurable control variate, the first-order
pooled variance depends on \(\sigma_j^2\) rather than
\(\sigma_j^2+\delta_j^2\). In the single-target case this makes Neyman allocation
reappear, while in the multi-target case the centered gap-aware proposal solves
the Gaussian characteristic-time problem. \cref{thm:lower-bound} therefore turns the
shared-allocation criterion from a procedural width proxy into a fundamental
Gaussian sample-complexity benchmark, while \cref{prop:ternary} and \cref{lem:klinf} supply the
corresponding bounded nonparametric limit for the reconstructed replay population.
\cref{prop:attainment} states sufficient conditions under which the Gaussian
first-order constant is attained. This sharpens the paper's central claim:
architecture does not merely change a convenient allocation formula; it changes
the relevant information geometry.

Third, selection and estimation should not be conflated. Pairwise point
estimates or fixed-time intervals may be adequate for descriptive
ranking, but declaring a best system after repeated monitoring is a
selection problem. The simultaneous anytime-valid construction used here
is intentionally conservative: on the joint coverage event, choosing the
current leader does not invalidate the pairwise confidence statements.
More efficient ranking-and-selection or best-arm methods can be layered
on top of the target-aware design framework when the number of systems
is large or selection efficiency is itself the primary objective.

The PromptEval replay provides an important counterpoint to a simplistic
``target changes the winner'' narrative. All 15 systems retain exactly
the same rank under macro, item-weighted, and STEM-emphasis targets, and
the leading pair has a positive mean difference in every MMLU subject.
Within the stratum-resolved architecture, item weighting reduces the
median from 33,085 under uniform sampling to 25,485 under
\(Q\)-proportional sampling, while the \(2/3\) range-width allocation
provides a further 5.9\% reduction to 23,985. But the larger result is
architectural within the local-CS comparator: a pooled target-sampling
PrPl-EB sequence requires 305 units at the median under batch-100
monitoring, with still smaller medians under the batch-10 sensitivity
analysis. Target-aware design therefore matters even when substantive
rankings are robust, but the dominant decision can be which uncertainty
guarantees must be retained, not only how observations are allocated.

Finally, the framework suggests a general software design principle for
sequential studies. A useful implementation should require the analyst
to declare the target weights and required inferential guarantees before
optimizing allocation, should preserve exploration or overlap when
adaptation is used, and should report stopping behavior alongside
terminal estimates. Such a separation would make the statistical meaning
of an adaptive design easier to audit across applications ranging from
surveys and experiments to simulation studies and model evaluation.

\section{Conclusion}\label{conclusion}

Sequential inference over heterogeneous strata should distinguish the
distribution defining the scientific target from the distribution used
to collect observations, and it should distinguish terminal estimation
from anytime-valid identification. For a target-weighted mean
difference, fixed-budget estimation yields the familiar Neyman
allocation. Under simultaneous stratum-resolved confidence sequences,
stopping depends instead on the geometry of the local time-uniform
boundaries, producing a boundary-rate allocation law whose root-n
variance-adaptive case has the 2/3 exponent. These are different
optimization problems, not competing universal prescriptions.

Target-aware sequential design therefore requires three choices that
should not be conflated: the target distribution, the inferential
architecture, and the sampling policy. For one prespecified target,
pooled direct inference can avoid the multiplicity and additive-width
cost of maintaining simultaneous local sequences. When stratum-resolved
guarantees or target flexibility are required, adaptive variance
tracking and the boundary-rate law provide principled allocation rules,
while predictable sampling preserves anytime validity. Simulations and
the PromptEval replay show that architecture choice can dominate fine
allocation tuning. The broader implication is simple: specify the target
and the guarantees first, then optimize where the next observation is
collected.

\clearpage
\section{References}\label{references}

Alshihry, Fatimah I., Tahani Coolen-Maturi, and Frank P. A. Coolen.
2026. ``Nonparametric Predictive Inference for Ranking and Selection.''
Journal of Statistical Theory and Practice 20: 93.
\url{https://doi.org/10.1007/s42519-026-00605-z}.

Angelopoulos, Anastasios N., Stephen Bates, Clara Fannjiang, Michael I.
Jordan, and Tijana Zrnic. 2023. ``Prediction-Powered Inference.''
Science 382 (6671): 669--674.
\url{https://doi.org/10.1126/science.adi6000}.

Audibert, Jean-Yves, Sébastien Bubeck, and Rémi Munos. 2010. ``Best Arm
Identification in Multi-Armed Bandits.'' In Proceedings of the 23rd
Annual Conference on Learning Theory, 41--53.

Berger, Roger L. 1982. ``Multiparameter Hypothesis Testing and Acceptance
Sampling.'' Technometrics 24 (4): 295--300.
\url{https://doi.org/10.1080/00401706.1982.10487790}.

Bommasani, Rishi, Percy Liang, and Tony Lee. 2023. ``Holistic Evaluation
of Language Models.'' Annals of the New York Academy of Sciences 1525
(1): 140--146. \url{https://doi.org/10.1111/nyas.15007}.

Darling, D. A., and Herbert Robbins. 1967. ``Confidence Sequences for
Mean, Variance, and Median.'' Proceedings of the National Academy of
Sciences of the United States of America 58 (1): 66--68.
\url{https://doi.org/10.1073/pnas.58.1.66}.

Étoré, Pierre, and Benjamin Jourdain. 2010. ``Adaptive Optimal
Allocation in Stratified Sampling Methods.'' Methodology and Computing
in Applied Probability 12 (3): 335--360.
\url{https://doi.org/10.1007/s11009-008-9108-0}.

Garivier, Aurélien, and Emilie Kaufmann. 2016. ``Optimal Best Arm
Identification with Fixed Confidence.'' In Proceedings of the 29th
Annual Conference on Learning Theory, Proceedings of Machine Learning
Research 49: 998--1027.

Hadad, Vitor, David A. Hirshberg, Ruohan Zhan, Stefan Wager, and Susan
Athey. 2021. ``Confidence Intervals for Policy Evaluation in Adaptive
Experiments.'' Proceedings of the National Academy of Sciences of the
United States of America 118 (15): e2014602118.
\url{https://doi.org/10.1073/pnas.2014602118}.

Hendrycks, Dan, Collin Burns, Steven Basart, Andy Zou, Mantas Mazeika,
Dawn Song, and Jacob Steinhardt. 2021. ``Measuring Massive Multitask
Language Understanding.'' In International Conference on Learning
Representations.

Hoeffding, Wassily. 1963. ``Probability Inequalities for Sums of Bounded
Random Variables.'' Journal of the American Statistical Association 58
(301): 13--30. \url{https://doi.org/10.1080/01621459.1963.10500830}.

Howard, Steven R., Aaditya Ramdas, Jon McAuliffe, and Jasjeet Sekhon.
2021. ``Time-Uniform, Nonparametric, Nonasymptotic Confidence
Sequences.'' The Annals of Statistics 49 (2): 1055--1080.
\url{https://doi.org/10.1214/20-AOS1991}.

Hsu, Chia-Yu, and Shubhanshu Shekhar. 2026. ``Efficient Sequential
Evaluation of Large Language Models.'' arXiv preprint arXiv:2607.17409.
\url{https://doi.org/10.48550/arXiv.2607.17409}.

Imbens, Guido, Lorenzo Masoero, Alexander Rakhlin, Thomas S. Richardson,
and Suhas Vijaykumar. 2026. ``Demonstration Experiments.'' arXiv preprint
arXiv:2603.06941. \url{https://doi.org/10.48550/arXiv.2603.06941}.

Jamieson, Kevin, Matthew Malloy, Robert Nowak, and Sébastien Bubeck.
2014. ``lil' UCB: An Optimal Exploration Algorithm for Multi-Armed
Bandits.'' Proceedings of Machine Learning Research 35: 423--439.

Johari, Ramesh, Pete Koomen, Leonid Pekelis, and David Walsh. 2021.
``Always Valid Inference: Continuous Monitoring of A/B Tests.''
Operations Research 70 (3): 1806--1821.
\url{https://doi.org/10.1287/opre.2021.2135}.

Karampatziakis, Nikos, Paul Mineiro, and Aaditya Ramdas. 2021.
``Off-Policy Confidence Sequences.'' In Proceedings of the 38th
International Conference on Machine Learning, Proceedings of Machine
Learning Research 139: 5301--5310.

Kaufmann, Emilie, Olivier Cappé, and Aurélien Garivier. 2016. ``On the
Complexity of Best-Arm Identification in Multi-Armed Bandit Models.''
Journal of Machine Learning Research 17 (1): 1--42.

Kim, Seong-Hee, and Barry L. Nelson. 2001. ``A Fully Sequential
Procedure for Indifference-Zone Selection in Simulation.'' ACM
Transactions on Modeling and Computer Simulation 11 (3): 251--273.
\url{https://doi.org/10.1145/502109.502111}.

Kish, Leslie. 1965. Survey Sampling. Wiley.

Lior, Gili, Eliya Habba, Shahar Levy, Avi Caciularu, and Gabriel
Stanovsky. 2025. ``ReliableEval: A Recipe for Stochastic LLM Evaluation
via Method of Moments.'' In Findings of the Association for
Computational Linguistics: EMNLP 2025, 11146--11153.
\url{https://doi.org/10.18653/v1/2025.findings-emnlp.594}.

Maurer, Andreas, and Massimiliano Pontil. 2009. ``Empirical Bernstein
Bounds and Sample-Variance Penalization.'' In Proceedings of the 22nd
Annual Conference on Learning Theory (COLT 2009).

Miller, Evan. 2024. ``Adding Error Bars to Evals: A Statistical Approach
to Language Model Evaluations.'' arXiv preprint arXiv:2411.00640.
\url{https://doi.org/10.48550/arXiv.2411.00640}.

Neuhof, Bitya, and Yuval Benjamini. 2026a. ``Quantifying Ranking
Uncertainty in LLM Benchmarks.'' arXiv preprint arXiv:2607.16259.
\url{https://doi.org/10.48550/arXiv.2607.16259}.

Neuhof, Bitya, and Yuval Benjamini. 2026b. ``Rank Intervals for
Leaderboards: A Hierarchical Framework for Model Evaluation.'' arXiv
preprint arXiv:2606.08679. \url{https://doi.org/10.48550/arXiv.2606.08679}.

Neyman, Jerzy. 1934. ``On the Two Different Aspects of the
Representative Method: The Method of Stratified Sampling and the Method
of Purposive Selection.'' Journal of the Royal Statistical Society 97
(4): 558--606. \url{https://doi.org/10.1111/j.2397-2335.1934.tb04184.x}.

O'Brien, Peter C., and Thomas R. Fleming. 1979. ``A Multiple Testing
Procedure for Clinical Trials.'' Biometrics 35 (3): 549--556.
\url{https://doi.org/10.2307/2530245}.

Paulson, Edward. 1964. ``A Sequential Procedure for Selecting the
Population with the Largest Mean from k Normal Populations.'' The Annals
of Mathematical Statistics 35 (1): 174--180.
\url{https://doi.org/10.1214/aoms/1177703739}.

Peng, Yijie, Chun-Hung Chen, Michael C. Fu, and Jian-Qiang Hu. 2016.
``Dynamic Sampling Allocation and Design Selection.'' INFORMS Journal on
Computing 28 (2): 195--208.
\url{https://doi.org/10.1287/ijoc.2015.0673}.

Pocock, Stuart J. 1977. ``Group Sequential Methods in the Design and
Analysis of Clinical Trials.'' Biometrika 64 (2): 191--199.
\url{https://doi.org/10.1093/biomet/64.2.191}.

Polo, Felipe Maia, Ronald Xu, Lucas Weber, Mírian Silva, Onkar Bhardwaj,
Leshem Choshen, Allysson Flavio Melo de Oliveira, Yuekai Sun, and
Mikhail Yurochkin. 2024. ``Efficient Multi-Prompt Evaluation of Large
Language Models.'' Advances in Neural Information Processing Systems 37:
22483--22512. \url{https://doi.org/10.52202/079017-0707}.

Ramdas, Aaditya, Peter Grünwald, Vladimir Vovk, and Glenn Shafer. 2023.
``Game-Theoretic Statistics and Safe Anytime-Valid Inference.''
Statistical Science 38 (4): 576--601.
\url{https://doi.org/10.1214/23-STS894}.

Robbins, Herbert. 1970. ``Statistical Methods Related to the Law of the
Iterated Logarithm.'' The Annals of Mathematical Statistics 41 (5):
1397--1409. \url{https://doi.org/10.1214/aoms/1177696786}.

Salehi, Mohammad, Mohammad Moradi, Jennifer A. Brown, and David R.
Smith. 2010. ``Efficient Estimators for Adaptive Stratified Sequential
Sampling.'' Journal of Statistical Computation and Simulation 80 (10):
1163--1179. \url{https://doi.org/10.1080/00949650903005664}.

Sandoval, Ricardo J., Ian Waudby-Smith, and Michael I. Jordan. 2026.
``Multi-Armed Sequential Hypothesis Testing by Betting.'' arXiv preprint
arXiv:2603.17925. \url{https://doi.org/10.48550/arXiv.2603.17925}.

Shafer, Glenn. 2021. ``Testing by Betting: A Strategy for Statistical
and Scientific Communication.'' Journal of the Royal Statistical Society
Series A: Statistics in Society 184 (2): 407--431.
\url{https://doi.org/10.1111/rssa.12647}.

Siska, Charlotte, Katerina Marazopoulou, Melissa Ailem, and James Bono.
2024. ``Examining the Robustness of LLM Evaluation to the Distributional
Assumptions of Benchmarks.'' In Proceedings of the 62nd Annual Meeting
of the Association for Computational Linguistics, 10406--10421.
\url{https://doi.org/10.18653/v1/2024.acl-long.560}.

Spertus, Jacob V., Mayuri Sridhar, and Philip B. Stark. 2024.
``Sequential Stratified Inference for the Mean.'' arXiv preprint
arXiv:2409.06680. \url{https://doi.org/10.48550/arXiv.2409.06680}.

Turner, Rosanne J., and Peter D. Grünwald. 2023. ``Safe Sequential
Testing and Effect Estimation in Stratified Count Data.'' Proceedings of
Machine Learning Research 206: 4880--4893.

Vovk, Vladimir, and Ruodu Wang. 2021. ``E-values: Calibration,
Combination, and Applications.'' The Annals of Statistics 49 (3):
1736--1754. \url{https://doi.org/10.1214/20-AOS2020}.

Wang, Hongjian, and Aaditya Ramdas. 2025. ``Anytime-Valid t-Tests and
Confidence Sequences for Gaussian Means with Unknown Variance.''
Sequential Analysis 44 (1): 56--110.
\url{https://doi.org/10.1080/07474946.2024.2428245}.

Waudby-Smith, Ian, Lili Wu, Aaditya Ramdas, Nikos Karampatziakis, and
Paul Mineiro. 2024a. ``Anytime-Valid Off-Policy Inference for Contextual
Bandits.'' ACM/IMS Journal of Data Science 1 (3): 1--42.
\url{https://doi.org/10.1145/3643693}.

Waudby-Smith, Ian, David Arbour, Ritwik Sinha, Edward H. Kennedy, and
Aaditya Ramdas. 2024b. ``Time-Uniform Central Limit Theory and Asymptotic
Confidence Sequences.'' The Annals of Statistics 52 (6): 2613--2640.
\url{https://doi.org/10.1214/24-AOS2408}.

Waudby-Smith, Ian, and Aaditya Ramdas. 2024. ``Estimating Means of
Bounded Random Variables by Betting.'' Journal of the Royal Statistical
Society Series B: Statistical Methodology 86 (1): 1--27.
\url{https://doi.org/10.1093/jrsssb/qkad009}.

Xu, Yang, Jiefu Zhang, Haixiang Sun, Zihan Zhou, Tianyu Cao, and Vaneet
Aggarwal. 2026. ``Towards Reliable LLM Evaluation: Correcting the
Winner's Curse in Adaptive Benchmarking.'' arXiv preprint
arXiv:2605.05973. \url{https://doi.org/10.48550/arXiv.2605.05973}.

\clearpage
\section*{Appendix}
\addcontentsline{toc}{section}{Appendix}
\appendix

\section{Proof of the estimation-optimal
allocation}\label{a.-proof-of-the-estimation-optimal-allocation}

Let

\[V\left( \mathbf{n} \right) = \sum_{j}^{}\frac{q_{j}^{2}\sigma_{j}^{2}}{n_{j}},\quad\quad\sum_{j}^{}c_{j}n_{j} = B.\]

The Lagrangian is

\[\mathcal{L} = \sum_{j}^{}\frac{q_{j}^{2}\sigma_{j}^{2}}{n_{j}} + \lambda\left( \sum_{j}^{}c_{j}n_{j} - B \right).\]

The first-order condition is

\[- \frac{q_{j}^{2}\sigma_{j}^{2}}{n_{j}^{2}} + \lambda c_{j} = 0,\]

so

\[n_{j} = \frac{q_{j}\sigma_{j}}{\sqrt{\lambda c_{j}}}.\]

Substitution into the budget constraint gives

\[\sqrt{\lambda} = \frac{\sum_{\ell}^{}q_{\ell}\sigma_{\ell}\sqrt{c_{\ell}}}{B},\]

which yields Equation~\eqref{eq:4.2}. Substitution back into \(V\) gives
Equation~\eqref{eq:4.3}. Convexity in each \(n_{j} > 0\) establishes the global
minimum. \(\square\)

\section{Proof of the boundary-rate allocation and
target-proportional
corollary}\label{b.-proof-of-the-boundary-rate-allocation-and-target-proportional-corollary}

For \cref{prop:boundary-rate}, minimize

\[{\widetilde{W}}_{\beta}\left( \mathbf{n} \right) = \sum_{j}^{}a_{j}n_{j}^{- \beta}\]

subject to \(\sum_{j}^{}c_{j}n_{j} = B\). The Lagrangian first-order
condition is

\[- \beta a_{j}n_{j}^{- (1 + \beta)} + \lambda c_{j} = 0,\]

so

\[n_{j} = \left( \frac{\beta a_{j}}{\lambda c_{j}} \right)^{1/(1 + \beta)}.\]

Let

\[S_{\beta} = \sum_{j}^{}a_{j}^{1/(1 + \beta)}c_{j}^{\beta/(1 + \beta)}.\]

The budget constraint gives
\((\beta/\lambda)^{1/(1 + \beta)} = B/S_{\beta}\), yielding
Equation~\eqref{eq:4.6}. Substitution into the objective gives

\[{\widetilde{W}}_{\beta,\min} = S_{\beta}^{1 + \beta}/B^{\beta}.\]

Because \(n \mapsto n^{- \beta}\) is strictly convex for \(\beta > 0\),
the solution is unique. Setting \(\beta = 1/2\) gives
Equations~\eqref{eq:4.8}--\eqref{eq:4.9}. \(\square\)

For \cref{cor:reversal}, under equal costs and a common range coefficient
\(r\), uniform allocation gives

\[W_{U} = \sum_{j}^{}\frac{q_{j}r}{\sqrt{B/J}} = \frac{r\sqrt{J}}{\sqrt{B}}.\]

Under target-proportional allocation,

\[W_{Q} = \sum_{j}^{}\frac{q_{j}r}{\sqrt{Bq_{j}}} = \frac{r}{\sqrt{B}}\sum_{j}^{}\sqrt{q_{j}}.\]

Cauchy--Schwarz implies
\(\sum_{j}^{}\sqrt{q_{j}} \leq \sqrt{J\sum_{j}^{}q_{j}} = \sqrt{J}\),
with equality exactly for uniform \(Q\). The fixed-budget variances
follow by substituting \(n_{j} = B/J\) and \(n_{j} = Bq_{j}\) into
Equation~\eqref{eq:4.1}, yielding Equation~\eqref{eq:4.12}. Comparing them gives
Equation~\eqref{eq:4.13}. \(\square\)

\section{Proof of the predictable local-time
lemma}\label{c.-proof-of-the-predictable-local-time-lemma}

Fix a stratum \(j\). By construction, \(A_{t}\) is known in
\(\mathcal{H}_{t}^{-}\) before \(D_{t}\) is revealed, and the
conditional mean restriction holds given \(\mathcal{H}_{t}^{-}\).
Consequently, the event that the \(r\)th visit to stratum \(j\) occurs
at time \(t\) is determined before the corresponding outcome is
observed. The visit times \(\tau_{j,r}\) are stopping times for the
natural filtration that includes the pre-outcome action at each time.

Define the local filtration

\[\mathcal{G}_{j,r} = \mathcal{H}_{\tau_{j,r}}\]

on \(\{\tau_{j,r} < \infty\}\), with the convention that the process and
filtration are held fixed after the final finite visit. A local
e-process can be written as a nonnegative product

\[M_{j,r} = \prod_{u = 1}^{r}E_{j,u},\]

where, under the null mean \(\delta_{j}\),

\[\mathbb{E}\left( E_{j,u} \mid \mathcal{G}_{j,u - 1} \right) \leq 1.\]

The conditional inequality remains valid because selection of the
\(u\)th stratum-\(j\) observation uses only pre-outcome information,
whereas the conditional mean model for that observation is assumed valid
given the full pre-outcome history \(\mathcal{H}_{t}^{-}\). Therefore

\[\mathbb{E}\left( M_{j,r} \mid \mathcal{G}_{j,r - 1} \right) \leq M_{j,r - 1}.\]

Hence \(\left( M_{j,r} \right)\) is a nonnegative supermartingale in
local time. In global time, \({\widetilde{M}}_{j,t} = M_{j,N_{j}(t)}\)
either remains unchanged or takes exactly one valid local-time update,
so

\[\mathbb{E}\left( {\widetilde{M}}_{j,t} \mid \mathcal{H}_{t - 1} \right) \leq {\widetilde{M}}_{j,t - 1}.\]

If the stratum is visited finitely often, holding the process constant
thereafter preserves the supermartingale property. Ville's inequality
then gives

\[\mathbb{P}\left( \exists t \geq 0:\ {\widetilde{M}}_{j,t} \geq 1/\alpha_{0} \right) \leq \alpha_{0}.\]

\(\square\)

\section{Proof of the target interval and best-system identification
results}\label{d.-proof-of-the-target-interval-and-best-system-identification-results}

For each stratum define the event

\[E_{j} = \{\delta_{j} \in C_{j,N_{j}(t)}\text{ for every }t\}.\]

By \cref{lem:local-time}, \(\mathbb{P}\left( E_{j}^{c} \right) \leq \alpha/J\).
Hence

\[\mathbb{P}\left( \underset{j = 1}{\bigcap^{J}}E_{j} \right) \geq 1 - \sum_{j}^{}\mathbb{P}\left( E_{j}^{c} \right) \geq 1 - \alpha.\]

On \(\cap_{j}E_{j}\), for every time \(t\),

\[L_{j,t} \leq \delta_{j} \leq U_{j,t}\]

for all \(j\). Multiplying by \(q_{j} \geq 0\) and summing yields

\[\sum_{j}^{}q_{j}L_{j,t} \leq \Delta_{Q} \leq \sum_{j}^{}q_{j}U_{j,t},\]

which proves \cref{thm:target-interval}.

For \(K\) systems, apply the same construction to each of the \(H\)
unordered pairs using local level \(\alpha/(HJ)\). A union bound over
the \(HJ\) pair-stratum processes yields a joint event of probability at
least \(1 - \alpha\) on which every pairwise target gap is covered at
every time. On that event, if system \(w\) satisfies Equation~\eqref{eq:7.1},
then \(\Delta_{wk,Q} > 0\) for all \(k \neq w\), so \(w\) is truly
uniquely best under the prespecified target distribution \(Q\).
Therefore the event in Equation~\eqref{eq:7.2} is contained in the complement of
the joint coverage event and has probability at most \(\alpha\). \(\square\)

\section{Proof of the adaptive optimal-allocation tracking theorem
and width
corollary}\label{e.-proof-of-the-adaptive-optimal-allocation-tracking-theorem-and-width-corollary}

Because \(\pi_{t + 1}(j) \geq \varepsilon_{t}\nu_{j}\) and
\(\sum_{t}^{}\varepsilon_{t} = \infty\), the conditional Borel--Cantelli
lemma implies that every target-relevant stratum is visited infinitely
often almost surely. Strong consistency of the within-stratum sample
variance along its local observation sequence then gives
\({\widehat{\sigma}}_{j,t} \rightarrow \sigma_{j}\) almost surely.
Continuity of \(x \mapsto \left( q_{j}x \right)^{2/3}\) on
\((0,\infty)\) yields \({\widehat{w}}_{j,t} \rightarrow w_{j}^{opt}\)
almost surely. Since \(\varepsilon_{t} \rightarrow 0\), also
\(\pi_{t + 1}(j) \rightarrow w_{j}^{opt}\).

For fixed \(j\), define
\(Y_{t,j} = \mathbf{1}\{ A_{t} = j\} - \pi_{t}(j)\). Then
\(\{ Y_{t,j}\}\) is a bounded martingale-difference sequence, and the
martingale strong law gives

\[t^{- 1}\sum_{s = 1}^{t}Y_{s,j} \rightarrow 0\quad\quad\text{almost surely}.\]

Cesàro convergence gives
\(t^{- 1}\sum_{s = 1}^{t}\pi_{s}(j) \rightarrow w_{j}^{opt}\). Combining
this with the preceding martingale convergence proves Equation~\eqref{eq:6.5}.

For \cref{cor:width-efficiency}, \cref{thm:tracking} gives
\(N_{j}(t) = tw_{j}^{opt}\{ 1 + o(1)\}\) almost surely. Because each
optimal share is positive and L is slowly varying,
\(L\{ N_{j}(t)\}/L(t) \rightarrow 1\). Hence

\[W_{t} = \kappa L(t)t^{- 1/2}\sum_{j}^{}q_{j}\sigma_{j}\left( w_{j}^{opt} \right)^{- 1/2}\{ 1 + o(1)\}.\]

The coefficient is exactly the minimum of
\(\sum_{j}^{}q_{j}\sigma_{j}w_{j}^{- 1/2}\) over simplex weights \(w\),
by \cref{prop:boundary-rate} with \(\beta = 1/2\). Thus the ratio to the optimal
first-order width converges to one.

\section{Proof of the robust target
results}\label{f.-proof-of-the-robust-target-results}

Introduce an epigraph variable \(z\) in Equation~\eqref{eq:5.1} and constraints
\(f_{m}\left( \mathbf{n} \right) \leq z\). The problem is convex and
satisfies Slater's condition. Let \(\lambda_{m} \geq 0\) be the KKT
multipliers for the \(M\) target constraints. Stationarity in \(z\)
gives \(\sum_{m}^{}\lambda_{m} = 1\), while complementary slackness
makes \(\lambda_{m} > 0\) only for active worst-case targets. The
weighted objective appearing in stationarity for \(n_{j}\) is

\[\sum_{m}^{}\lambda_{m}f_{m}\left( \mathbf{n} \right) = \sum_{j}^{}\left( \sum_{m}^{}\lambda_{m}q_{j}^{(m)} \right)r_{j}n_{j}^{- \beta} = \sum_{j}^{}{\bar{q}}_{j}r_{j}n_{j}^{- \beta}.\]

Applying \cref{prop:boundary-rate} to this weighted objective yields
Equation~\eqref{eq:5.2}. Strong duality establishes optimality. \(\square\)

For \cref{prop:uniform-targetset}, on the joint local coverage event, for every
\(j,t\) we have \(L_{j,t} \leq \delta_{j} \leq U_{j,t}\). Thus for every
\(Q \in \mathcal{Q}\),

\[\sum_{j}^{}q_{j}L_{j,t} \leq \Delta_{Q} \leq \sum_{j}^{}q_{j}U_{j,t}.\]

Both sides are linear in \(Q\). Their extrema over a convex hull are
attained at vertices, giving exactly Equation~\eqref{eq:6.8}. The probability of
this simultaneous event is already at least \(1 - \alpha\); no
additional union bound over \(M\) is taken. \(\square\)

\section{Proofs for target-set pooled inference and shared proposals}\label{g.-targetset-proofs}

For \cref{prop:targetset-pooled}, let
\(E_m=\{\Delta_{Q^{(m)}}\in C_t^{(m)}\text{ for every }t\}\).
By anytime-validity,
\(\mathbb P(E_m^c)\leq\alpha_m\), so the union bound gives
\[
\mathbb P\left(\bigcap_{m=1}^M E_m\right)
\geq 1-\sum_{m=1}^M\alpha_m\geq 1-\alpha.
\]
On this event, for any
\(Q=\sum_m\omega_mQ^{(m)}\) with \(\omega_m\geq0\) and
\(\sum_m\omega_m=1\), linearity gives
\(\Delta_Q=\sum_m\omega_m\Delta_{Q^{(m)}}\). Multiplying the
vertexwise bounds by \(\omega_m\) and summing proves
\[
\sum_m\omega_mL_t^{(m)}\leq\Delta_Q\leq
\sum_m\omega_mU_t^{(m)}.
\]
If every lower endpoint is positive, then every convex combination is
positive; the negative statement is analogous. \(\square\)

For \cref{prop:minimax-proposal}, write
\(a_{mj}=(q_j^{(m)})^2m_{2j}\) and introduce an epigraph variable
\(z\). The primal problem is
\[
\min_{\pi,z} z
\quad\text{subject to}\quad
\sum_j\frac{a_{mj}}{\pi_j}\leq z\ (m=1,\ldots,M),
\quad \pi_j>0,\quad\sum_j\pi_j=1.
\]
It is convex and satisfies Slater's condition on the active strata. Let
\(\lambda_m\geq0\) be the multipliers of the \(M\) target
constraints. Minimization over \(z\) requires
\(\sum_m\lambda_m=1\). For fixed
\(\lambda\in\Delta_M\), define
\[
A_j(\lambda)=m_{2j}\sum_{m=1}^M\lambda_m(q_j^{(m)})^2.
\]
The inner proposal problem is
\(\min_{\pi\in\Delta_J}\sum_jA_j(\lambda)/\pi_j\). By
Cauchy--Schwarz,
\[
\left(\sum_j\frac{A_j}{\pi_j}\right)
\left(\sum_j\pi_j\right)
\geq\left(\sum_j\sqrt{A_j}\right)^2,
\]
with equality exactly when \(\pi_j\propto\sqrt{A_j}\). Strong
duality therefore yields a maximizing least-favourable
\(\lambda^*\in\Delta_M\) and Equation~\eqref{eq:6.13}. Complementary
slackness implies that positive components of \(\lambda^*\) occur
only for target constraints active at the minimax optimum. \(\square\)

For \cref{cor:waterfill}, the cap
\(B_{\mathcal Q}(\pi)\leq C\) is equivalent to
\(\pi_j\geq\ell_j(C)=\max_mq_j^{(m)}/C\). Such a proposal exists
if and only if \(\sum_j\ell_j(C)\leq1\), which is Equation~\eqref{eq:6.15}.
For fixed dual weights \(\lambda\), minimize
\(\sum_jA_j(\lambda)/\pi_j\) subject to these lower bounds and the
simplex constraint. The KKT conditions give
\[
-\frac{A_j}{\pi_j^2}+\nu-\mu_j=0,
\qquad \mu_j\geq0,
\qquad \mu_j(\pi_j-\ell_j)=0.
\]
Hence an inactive lower bound has
\(\pi_j=\sqrt{A_j/\nu}=\kappa\sqrt{A_j}\), whereas an active
one has \(\pi_j=\ell_j\). Thus
\(\pi_j=\max\{\ell_j,\kappa\sqrt{A_j}\}\), with the unique
\(\kappa>0\) chosen to make the probabilities sum to one. Maximizing
the resulting dual objective over \(\lambda\in\Delta_M\) gives
Equation~\eqref{eq:6.16}. For \(M=1\),
\(\sqrt{A_j}=q_j\sqrt{m_{2j}}\), yielding the stated single-target
water-filling form. \(\square\)

\paragraph{Proof of \cref{prop:min-range}.}
Let $r_j=\max_m q_j^{(m)}$. If $B_{\mathcal Q}(\pi)\leq C$, then
$r_j\leq C\pi_j$ for every $j$, hence
$\sum_jr_j\leq C\sum_j\pi_j=C$. Therefore
$B_{\mathcal Q}(\pi)\geq\sum_jr_j=C_{\mathcal Q}$ for every proposal.
At $\pi_j=r_j/C_{\mathcal Q}$ the probabilities sum to one and
$r_j/\pi_j=C_{\mathcal Q}$ on every active stratum, proving attainability.
Because each target distribution sums to one,
$1\leq\sum_j\max_mq_j^{(m)}\leq\sum_{m,j}q_j^{(m)}=M$.
If $C_{\mathcal Q}=1$, then every $Q^{(m)}$ is coordinatewise bounded by
the vector $r=(r_j)$ and both have total mass one, so every target must equal
$r$; the converse is immediate. $\square$

\paragraph{Intersection--union argument.}
Because $Q\mapsto\Delta_Q$ is linear,
$\min_{Q\in\mathcal Q}\Delta_Q=\min_m\Delta_{Q^{(m)}}$.
Under $H_{0,+}$ there is therefore at least one vertex $m^*$ with
$\Delta_{Q^{(m^*)}}\leq0$. If $\tau_+<\infty$, then at time $\tau_+$,
$L_{\tau_+}^{(m^*)}>0\geq\Delta_{Q^{(m^*)}}$, so the anytime lower
confidence sequence for vertex $m^*$ has failed. Hence
\[
\Pr_{P}(\tau_+<\infty)\leq
\Pr_P\{\exists t:L_t^{(m^*)}>\Delta_{Q^{(m^*)}}\}\leq\alpha.
\]
No independence assumption and no union bound over $m$ is used. The upper-bound
argument is symmetric. If both directional procedures are run at level
$\alpha/2$, a union bound over the two possible wrong directional declarations
gives total error at most $\alpha$. $\square$

\paragraph{Proof of \cref{prop:gap-aware}.}
Let
$a_{mj}=(q_j^{(m)})^2m_{2j}/g_m^2$. Equation~\eqref{eq:6.17} is exactly the convex
program $\min_{\pi\in\Delta_J}\max_m\sum_ja_{mj}/\pi_j$.
Repeating the epigraph duality argument used for \cref{prop:minimax-proposal} gives, for
$\lambda\in\Delta_M$,
\[
A_j(\lambda)=m_{2j}\sum_m\lambda_m(q_j^{(m)})^2/g_m^2.
\]
Cauchy--Schwarz yields
$\inf_{\pi\in\Delta_J}\sum_jA_j(\lambda)/\pi_j
=(\sum_j\sqrt{A_j(\lambda)})^2$, attained at
$\pi_j\propto\sqrt{A_j(\lambda)}$. Strong duality gives Equations~\eqref{eq:6.18}--\eqref{eq:6.19},
and complementary slackness gives the support statement for
$\lambda^\dagger$. Since
$V_m(\pi)/g_m^2=S_m(\pi)/g_m^2-1$, the same proposal minimizes the worst
first-order variance-to-gap-squared stopping proxy. The range-capped form
follows from the KKT proof of \cref{cor:waterfill} with these gap-scaled coefficients.
$\square$

\paragraph{Proof of \cref{thm:gap-tracking}.}
The exploration condition implies that every target-relevant stratum is visited
infinitely often almost surely. Bounded i.i.d. sampling within each stratum then
gives $\widehat\delta_{j,t}\to\delta_j$ and
$\widehat m_{2j,t}\to m_{2j}$ almost surely. Because every $g_m>0$ and
$\gamma_t\downarrow0$, eventually
$\widehat g_{m,t}\to g_m$ for all $m$. The plug-in objective functions in
Equation~\eqref{eq:6.17} therefore converge to the population objective on every compact
subset of the active simplex. Uniqueness of $\pi^\dagger$ and standard argmin
continuity imply $\widehat\pi_t^\dagger\to\pi^\dagger$ almost surely.
Since $\varepsilon_t\to0$,
$\widetilde\pi_t\to\pi^\dagger$. Finally, for each $j$,
$1\{A_t=j\}-\widetilde\pi_t(j)$ is a bounded martingale-difference sequence;
the martingale strong law and Cesaro convergence give
$N_j(t)/t\to\pi_j^\dagger$ almost surely. $\square$

\paragraph{Proof of \cref{cor:tv}.}
For two probability vectors $q^{(1)}$ and $q^{(2)}$,
$\max(a,b)=(a+b+|a-b|)/2$. Therefore
\[
C_{\{Q^{(1)},Q^{(2)}\}}
=\sum_j\max\{q_j^{(1)},q_j^{(2)}\}
=1+\frac12\sum_j|q_j^{(1)}-q_j^{(2)}|,
\]
which is exactly $1+\operatorname{TV}(Q^{(1)},Q^{(2)})$. $\square$

\paragraph{Proof of \cref{prop:centering}.}
Condition on $\mathcal H_{t-1}$. The centering terms are predictable and
$A_t$ is drawn from $\pi_t$ before the current outcome is observed. Hence
\[
\begin{aligned}
\mathbb E(Y_t^{(m)}\mid\mathcal H_{t-1})
&=\sum_jq_j^{(m)}\widehat\delta_{j,t-1}
+\sum_j\pi_t(j)\frac{q_j^{(m)}}{\pi_t(j)}
(\delta_j-\widehat\delta_{j,t-1})\\
&=\sum_jq_j^{(m)}\delta_j=\Delta_{Q^{(m)}}.
\end{aligned}
\]
The random part is
$X_t=q_{A_t}^{(m)}\{D_t-\widehat\delta_{A_t,t-1}\}/\pi_t(A_t)$.
Its conditional mean is $r_{m,t-1}$ and
\[
\mathbb E(X_t^2\mid\mathcal H_{t-1})
=\sum_j\frac{(q_j^{(m)})^2}{\pi_t(j)}
\{\sigma_j^2+(\delta_j-\widehat\delta_{j,t-1})^2\}.
\]
Subtracting $r_{m,t-1}^2$ proves Equation~\eqref{eq:6.22}. Consistency of the
predictable centering estimates and convergence of the proposal give
Equation~\eqref{eq:6.23} by termwise convergence. For one target, Cauchy--Schwarz gives
\[
\left(\sum_j\frac{q_j^2\sigma_j^2}{\pi_j}\right)
\left(\sum_j\pi_j\right)
\geq\left(\sum_jq_j\sigma_j\right)^2,
\]
with equality exactly when $\pi_j\propto q_j\sigma_j$, proving
Equation~\eqref{eq:6.24}. $\square$

\paragraph{Proof of \cref{thm:lower-bound}.}
Let $E$ denote the event that the procedure declares robust positivity. The
bandit change-of-measure inequality implies that for every null parameter
$\delta'\in\mathcal H_{0,+}$,
\[
\sum_j\mathbb E_\delta N_j(\tau)\,
\mathrm{KL}\!\left(N(\delta_j,\sigma_j^2),
N(\delta'_j,\sigma_j^2)\right)
\geq \operatorname{kl}\{\mathbb P_\delta(E),
\mathbb P_{\delta'}(E)\}
\geq\operatorname{kl}(1-\beta,\alpha).
\]
Write $T=\mathbb E_\delta\tau$ and, when $T>0$,
$w_j=\mathbb E_\delta N_j(\tau)/T$. For Gaussian observations the left side
becomes
\[
T\sum_jw_j\frac{(\delta_j-\delta'_j)^2}{2\sigma_j^2}.
\]
The robust null is a union over target vertices, so its closest point is obtained
by minimizing separately over each half-space
$(Q^{(m)})^\top\delta'\leq0$. Because the true gap $g_m$ is positive, the
constraint is active at the minimizer. Put $h_j=\delta_j-\delta'_j$; then
$\sum_jq_j^{(m)}h_j=g_m$. Lagrange multipliers, or Cauchy--Schwarz, give
\[
\inf_{\sum_jq_j^{(m)}h_j=g_m}
\sum_jw_j\frac{h_j^2}{2\sigma_j^2}
=\frac{g_m^2}{2\sum_j(q_j^{(m)})^2\sigma_j^2/w_j}
=\frac{g_m^2}{2V_m(w)}.
\]
Therefore
$T\min_m g_m^2/(2V_m(w))\geq\operatorname{kl}(1-\beta,\alpha)$.
The minimum information rate for this $w$ is no larger than
$\Gamma^*$, proving Equation~\eqref{eq:6.26}. If $\mathbb P_\delta(E)=1$, the binary
relative entropy equals $\operatorname{kl}(1,\alpha)=\log(1/\alpha)$, which
gives Equation~\eqref{eq:6.27}. $\square$

\paragraph{Proof of \cref{cor:info-optimal}.}
Taking reciprocals in the positive max--min problem gives
$1/\Gamma^*=2D^*$ with
$D^*=\min_w\max_m V_m(w)/g_m^2$. This is the same convex program as
\cref{prop:gap-aware} with coefficients
$a_{mj}=(q_j^{(m)})^2\sigma_j^2/g_m^2$. Repeating the epigraph duality and
Cauchy--Schwarz argument yields Equations~\eqref{eq:6.28}--\eqref{eq:6.29}. For $M=1$ the dual
weight is one, so $w_j^*\propto q_j\sigma_j$ and
$D^*=(\sum_jq_j\sigma_j)^2/\Delta_Q^2$, proving Equation~\eqref{eq:6.30}. $\square$

\paragraph{Proof of \cref{prop:attainment}.}
Let $c_m=2V_m(w^*)/g_m^2$ and
$c^*=\max_m c_m=1/\Gamma^*$. By the assumed uniform consistency and boundary
expansion, for every fixed $\eta>0$, uniformly over target vertices and times
$t$ on the $\log(1/\alpha)$ scale,
\[
L_{m,t}=g_m-
\sqrt{\frac{2V_m(w^*)\log(1/\alpha)}{t}}+o_{\mathbb P}(1).
\]
At $t=(c^*+\eta)\log(1/\alpha)$, the leading term is strictly positive for
every $m$, so $\Pr\{\tau_\alpha>t\}\to0$. Conversely, choose an active vertex
$m^*$ with $c_{m^*}=c^*$. At
$t=(c^*-\eta)\log(1/\alpha)$, the leading term for $m^*$ is strictly negative,
so $\Pr\{\tau_\alpha\leq t\}\to0$. These two inequalities imply
$\tau_\alpha/\log(1/\alpha)\to c^*=1/\Gamma^*$ in probability. Uniform
integrability, when assumed, upgrades this convergence to convergence of the
expectations. $\square$

\paragraph{Proof of \cref{prop:ternary}.}
Fix a target vertex \(m\) and an allocation \(w\) with positive coordinates.
Write the support points as \(x\in\{-1,0,1\}\), the true stratum probabilities
as \(p_{jx}\), and an alternative as \(r_{jx}\).  Minimize
\(\sum_jw_j\sum_xp_{jx}\log(p_{jx}/r_{jx})\) subject to the stratum
normalization constraints and
\(\sum_jq_j^{(m)}\sum_xxr_{jx}\leq0\).  Because the true target gap is
positive, the target-mean constraint is active at the I-projection.  The KKT
first-order condition gives
\[
-\frac{w_jp_{jx}}{r_{jx}}+a_j+\lambda q_j^{(m)}x=0,
\]
so, after writing \(c_j=a_j/w_j\),
\[
r_{jx}=\frac{p_{jx}}{c_j+(\lambda q_j^{(m)}/w_j)x}.
\]
For fixed \(\lambda\), strict convexity of reverse KL in the alternative and
normalization determine a unique admissible \(c_j\); the common multiplier
\(\lambda\) is then chosen so that the target mean equals zero.  The adaptive
change-of-measure inequality applied to any null alternative in the ternary
model gives
\(E_P\tau\sum_jw_j\mathrm{KL}(P_j\Vert R_j)\geq
\operatorname{kl}(1-\beta,\alpha)\).  Minimizing over the ternary null and
maximizing the resulting information rate over \(w\) proves Equation~\eqref{eq:6.34}.
A procedure valid over all \([-1,1]\)-supported stratum laws is in particular
valid over the ternary model, so the bound applies to it.
\(\square\)

\paragraph{Proof of \cref{lem:klinf}.}
The ternary infimum is taken over a subset of the laws on \([-1,1]\), so it is
at least the unrestricted infimum. For the reverse inequality, take any
admissible family \(\{R_j\}\) of laws on \([-1,1]\) with
\(P_j\ll R_j\) and \(\sum_jq_j^{(m)}E_{R_j}D\leq0\). Let \(s_j\) be the
mass that \(R_j\) places on \([-1,1]\setminus\{-1,0,1\}\), and define
\(R_j'\) on \(\{-1,0,1\}\) by keeping \(R_j\) on \(\{-1,0,1\}\) and adding
the mass \(s_j\) to the point \(-1\). Then \(R_j'(x)\geq R_j(x)\) for every
\(x\) in the support of \(P_j\), and \(\mathrm{KL}(P_j\Vert R)=\sum_x
p_{jx}\log(p_{jx}/r_x)\) is nonincreasing in each \(r_x\) on that support, so
\(\mathrm{KL}(P_j\Vert R_j')\leq\mathrm{KL}(P_j\Vert R_j)\). Moving mass to
the smallest support point cannot increase the mean, so
\(E_{R_j'}D\leq E_{R_j}D\) and the target constraint still holds. Hence every
admissible law on \([-1,1]\) is dominated by an admissible ternary law with no
larger weighted KL divergence, and the two infima coincide.
\(\square\)

\section{Reproducibility}\label{h.-reproducibility}

The supplementary materials contain: (i) the base-R reproduction script
for Tables~1--3, Table~7, and the optional-stopping experiment; (ii)
reproduce\_variance\_adaptive\_upgrade\_fairbatch.R for Tables~4--6,
with the R 4.3.3 console output for the reported results; (iii) the
rate-limit-aware PromptEval retrieval/replay script with compact CSV/RDS
outputs for Tables~8 and~9 and the subject-level diagnostics; and (iv)
reproduce\_direct\_pooled\_cs.R for the pooled target and
importance-weighted baselines; (v)
\path{reproduce_targetset_replay.py} for the minimax, gap-aware, and
range-capped shared proposals, the simultaneous target-set replay in Table~11,
the robust-positive replay in Table~12, and the matched one-sided single-target
baseline; (vi) \path{reproduce_information_bound_v2.py} for the Gaussian
characteristic-time benchmark and information-optimal shared proposal in
Table~13; and (vii) \path{reproduce_ternary_information_bound.py} for the
ternary (bounded nonparametric) benchmark in Table~13.
The pooled scripts reconstruct each exact ternary subject distribution
from \(\delta_{j}\) and \(\sigma_{j}\). The supplied batch-100 output
is used for Table~10, while the batch-10 output provides the sensitivity
results reported in Sections~8.5 and~10.3. The Python scripts use
NumPy, pandas, and SciPy; the target-set script writes
\path{targetset_pooled_results_batch10.csv},
\path{targetset_robust_positive_results_one_sided_batch10.csv},
\path{single_target_one_sided_batch10.csv}, and the proposal files.

\end{document}